\documentclass[twocolumn,showpacs,prc,aps,floatfix,hidelinks]{revtex4-2}

\usepackage{bm}
\usepackage{braket}
\usepackage{amssymb}
\usepackage{graphicx}
\usepackage{dcolumn}
\usepackage{amsmath}
\usepackage{amsfonts}
\usepackage{xcolor}
\usepackage{hyperref}
\usepackage{multirow}
\usepackage{tabularx}
\usepackage{physics}
\usepackage[version=4]{mhchem}
\usepackage{caption}
\usepackage{subcaption}
\usepackage{mathtools}

\usepackage{slashed}

\begin{document}

\title{\texorpdfstring{\textit{Ab initio} investigation of the $\ce{^7Li}(p,e^+e^-)\ce{^8Be}$ process and the X17 boson}{Ab initio investigation of the 7Li(p,e+e-)8Be process and the X17 boson}}
\author{P. Gysbers$^{1,2,3}$, P. Navr\'atil$^{1,4}$, K.~Kravvaris$^5$, G.~Hupin$^6$, S.~Quaglioni$^5$} 
\affiliation{${}^1$TRIUMF, Vancouver, British Columbia, V6T 2A3, Canada}
\affiliation{${}^2$Department of Physics and Astronomy, University of British Columbia, Vancouver, British Columbia, V6T 1Z1, Canada}
\affiliation{${}^3$Facility for Rare Isotope Beams, Michigan State University, East Lansing, MI 48824, USA}
\affiliation{${}^4$University of Victoria, 3800 Finnerty Road, Victoria, British Columbia V8P 5C2, Canada}
\affiliation{${}^5$Lawrence Livermore National Laboratory, P.O. Box 808, L-414, Livermore, CA 94551, USA}
\affiliation{${}^6$Université Paris-Saclay, CNRS/IN2P3, IJCLab, 91405 Orsay, France}

\date{\today}

\begin{abstract}
Observations of anomalies in the electron-positron angular correlations in high-energy decays in $^4$He, $^8$Be, and $^{12}$C have been reported recently by the ATOMKI collaboration. These could be explained by the creation and subsequent decay of a new boson with a mass of ${\approx}17$~MeV. Theoretical understanding of pair creation in the proton capture reactions used in these experiments is important for the interpretation of the anomalies. We apply the {\it ab initio} No-Core Shell Model with Continuum (NCSMC) to the proton capture on $^7$Li. The NCSMC describes both bound and unbound states in light nuclei in a unified way with chiral two- and three-nucleon interactions as the only input. We investigate the structure of $^8$Be, the $p+^7$Li elastic scattering, the $^7$Li($p,\gamma$)$^8$Be cross section and the internal pair creation $^7$Li($p,e^+ e^-$)$^8$Be. We discuss the impact of a proper treatment of the initial scattering state on the electron-positron angular correlation spectrum and compare our results to available ATOMKI data sets. Finally, we calculate $^7$Li($p,X$)$^8$Be cross sections for several proposed models of the hypothetical X17 particle.
\end{abstract}

\maketitle

\section{Introduction}

Motivated by Weinberg's~\cite{Weinberg1978} and Wilczek's~\cite{Wilczek1978} predictions of a new light boson particle, the axion, that could resolve the strong CP (charge conjugation parity symmetry) problem of quantum chromodynamics (QCD) and by the suggestion by Donnelly {\it et al.}~\cite{Donnelly1978} to study the angular correlation of the $e^+ e^-$ pairs created in $1^+\rightarrow 0^+$ nuclear transitions as a signature for the decay of the axion, experiments have been launched to detect the new particle. However, no such particle was found in a ${\approx} 1$~MeV/c$^2$ mass range~\cite{Savage1986,Savage1988}. Later, an observation of a ${\approx} 9$ MeV particle was claimed~\cite{FWNdeBoer_2001}, although it has not been confirmed. 
Similarly, motivated by the standard model Higgs boson prediction, unsuccessful searches have been performed for light scalar boson particles in nuclear decays with large energy release~\cite{Freedman1984,Savage1988}. Later, experimental searches for light vector or pseudoscalar (axion-like) particles had nominally excluded particles with such quantum numbers in the mass range of tens of MeV~\cite{NA482015,Dobrich2016}.

Recently, the ATOMKI collaboration measured the angular correlation of the $e^+ e^-$ pairs created following the proton capture reaction on $^7$Li. They reported an anomaly at ${\approx}140^\circ$ in the transition from the second $1^+$ resonance above the $p{+}^7$Li threshold, at the $^8$Be excitation energy of 18.15 MeV, which is predominantly isoscalar, and interpreted it as the decay of a new boson, called X17, of mass ${\approx} 17$ MeV/c$^2$~\cite{Krasznahorkay2016,Firak2020}. No such anomaly was seen in the dominantly isovector transition from the first $1^+$ resonance at the $^8$Be excitation energy of 17.64 MeV. 

The ATOMKI collaboration then performed a similar measurement for the $e^+ e^-$ internal pair conversion in high energy transitions to the $^4$He ground state following the capture of a proton from a triton target. They reported again an anomaly in the internal pair angular correlations that could be explained by a decay of a boson with about the same mass as that seen in the $^8$Be decay~\cite{PhysRevC.104.044003}. However, the transitions appeared to be of the $E1$ (vector) character, i.e., it could not be interpreted as an axion decay. 

A later new measurement of the proton capture on $^7$Li, exploring also  energies between the $1^+$ resonances and slightly above the 18.15 MeV resonance, reported an $e^+ e^-$ pair angular correlations anomaly in the off-resonance transitions implying an $E1$ character of the decay~\cite{Sas:2022pgm}. Very recently, the ATOMKI collaboration reported the X17 anomaly also in the decay of the $^8$Be giant dipole resonance in transitions to both the ground state and the broad first excited state~\cite{Krasznahorkay2023}.

In their latest experiment, the ATOMKI collaboration investigated the proton capture on $^{11}$B at energies covering the broad 17.23 MeV $1^-$ resonance in $^{12}$C reporting once again an anomaly in the $e^+ e^-$ internal pair conversion angular correlations from the transitions to the ground state consistent with a ${\approx} 17$ MeV boson, the X17 particle, of a vector character~\cite{Krasznahorkay2022}.

This X17 anomaly triggered many theoretical interpretations on the particle physics side, exploring, e.g., signatures of axion-like particles, vector or axial vector bosons, and dark photons~\cite{Kozaczuk2017,Feng2016,Feng2017,Feng2020,Ellwanger2016,Alves2021,Backens2022,aleksejevs2021standard,universe10040173,kubarovsky2022quantum,Denton2023,Barducci2023,Alves2023}. However, other possible explanations need to be excluded first, including possible issues with the observation or the interpretation of the data. On the observation side, several new experiments have been proposed and initiated to provide an independent verification of the anomaly, with data collection already completed for some of them~\cite{Azuelos_2022,Darm__2022,Cline_2022,NewJedi2023,mommers2023x17}. At the same time, the CERN experiment NA64 eliminated much (but not all) of the allowed parameter space for a vector X17 boson~\cite{NA642020} with follow-up experiments proposed at the CERN SPS~\cite{Depero2020}. 
On the interpretation side, it is worthwhile to investigate the pair production processes considering the complex nuclear structure and reaction effects~\cite{ZHANG2017159,Viviani2022,Hayes2022}. 

The nuclear transition form factor as a possible origin of the anomaly was investigated in Ref.~\cite{ZHANG2017159} but it was found that the required form factor is unrealistic for the $^8$Be nucleus. A detailed {\it ab initio} investigation of the internal pair conversion and creation and decay of various hypothetical bosons in the proton capture on $^3$H has been performed using the hyperspherical harmonics method and realistic nucleon-nucleon (NN) and three-nucleon (3N) forces in Ref.~\cite{Viviani2022}. Overall, these calculations were able to reasonably reproduce the ATOMKI $p+^3$H data although they called for more accurate measurements of the pair-production cross sections by performing experimental studies in a wider range of energy and $e^+ e^-$ angles. Angular correlations in the $e^+ e^-$ decay of the 18.15 MeV excited state in $^8$Be were investigated in Ref.~\cite{Hayes2022}. The available cross sections and angular distributions for the $^7$Li($p,\gamma$) proton capture reaction were examined through an R-matrix analysis that established the dominance of the $E1$ and $M1$ multipoles in the transition. The resulting analysis indicated that the ATOMKI measured $e^+ e^-$ angular correlations fall off too rapidly with angle, falling below the R-matrix expectations at angles greater than 100$^\circ$.

In this work, we present an in-depth study of the $p{+}^7$Li capture reaction within the {\it ab initio} no-core shell model with continuum (NCSMC) approach that allows us to describe simultaneously the structure of $^8$Be, the $^7$Li($p,p$) proton elastic scattering, 
the radiative capture $^7$Li($p,\gamma$)$^8$Be, the internal pair conversion $^7$Li($p,e^+ e^-$)$^8$Be, as well as the X17 boson production $^7$Li($p,X$)$^8$Be and decay for a variety of candidates for the hypothetical boson. It should be noted that within the NCSMC formalism, contrary to other methods that have been applied to calculate the $^8$Be decays so far, we are able to compute both resonant (emission from a discrete nuclear excited state) and non-resonant $e^+e^-$ pair production in pertinent nuclear processes (in this paper for the production of $^8$Be) with realistic isospin-breaking effects, mediated by a virtual photon or by X17. Our formalism for the $e^+ e^-$ internal pair conversion is consistent with that of Ref.~\cite{Viviani2022}, which is somewhat more general than that of Ref.~\cite{Hayes2022}. We compare our results to the 2016/2019 ATOMKI data~\cite{Krasznahorkay2016,Firak2020} as well as to their latest measurements performed at the two $1^+$ resonances and at two energies between the resonances~\cite{Sas:2022pgm}. The rest of the paper is organized as follows. In Sect.~\ref{theory}, we briefly review the NCSMC method, the description of radiative capture, and the internal $e^+e^-$ pair production theory. We present our results in Sect.~\ref{results}, and conclude in Sect.~\ref{concl}. The nuclear transition matrix elements of electromagnetic multipole operators are briefly reviewed in Appendix~\ref{sec:multipole_operators}. The details of the derivation of the radiative capture and the internal $e^+e^-$ pair production cross sections are presented in Appendix~\ref{sec:radcap_appendix} and \ref{sec:pairprod_radcap_appendix}, respectively.

\section{Theory}\label{theory}

\subsection{NCSMC formalism}\label{NCSMC_theory}

As we aim at describing the processes in the $^8$Be composite system in-between the $p{+}^7$Li and $n{+}^7$Be thresholds, we must consider explicitly the $p{+}^7$Li and $n{+}^7$Be clusters. This is more general than what was done in past NCSMC applications, where the inter cluster dynamics was described using the isospin as a good quantum number~\cite{Baroni2013L,Baroni2013,physcripnavratil}. The ansatz for the NCSMC wave functions is a generalized cluster expansion
\begin{align}
   \Ket{\Psi^{J^\pi}_A} = &\sum_\lambda c_\lambda^{J^\pi}\Ket{A \lambda J^\pi} \nonumber\\
   &+ \sum_\nu\int drr^2\frac{\gamma_{\nu}^{J^\pi}(r)}{r}\hat{\mathcal{A}}_\nu\Ket{\Phi_{\nu r}^{J^\pi}}.
   \label{eq:ncsmc}
\end{align}
The first term is an expansion over no-core shell model (NCSM)~\cite{PhysRevLett.84.5728,PhysRevC.62.054311,Navratil2009,BARRETT2013131} eigenstates of the aggregate system $\Ket{A \lambda J^\pi}$ (here $^{8}$Be, $A{=8}$) calculated in a many-body harmonic oscillator (HO) basis. The second term is an expansion over microscopic cluster channels
$\hat{\mathcal{A}}_\nu\Ket{\Phi_{\nu r}^{J^\pi}}$ which describe the clusters ($^{7}$Li+$p$ and $^{7}$Be+$n$) in relative motion:
\begin{align}
    \ket{\Phi^{J^\pi}_{\nu r}} = &\Big[ \big( \ket{(A-1) \, \alpha I^{\pi_t}} \ket{N \, \tfrac12^{\texttt{+}}} \big)^{(s)}\;
     Y_\ell(\hat{r}_{A-1,1}) \Big]^{(J^{\pi})}  \nonumber \\
    &\times\,\frac{\delta(r{-}r_{A-1,1})}{rr_{A-1,1}} \; , 
    \label{eq:rgm_state}
\end{align}
where 
$\ket{(A-1) \, \alpha I^{\pi_t}}$ and $\ket{N \, \tfrac12^{\texttt{+}}}$ are the eigenstates of the target ($^7$Li or $^7$Be) and the single-nucleon projectile $N$ ($p$ or $n$), respectively. The cluster channels enable the description of scattering states as well as bound states including extended (halo) states in the NCSMC. The $^{7}$Li and $^7$Be eigenstates (calculated within the NCSM) have angular momentum $I$, parity $\pi_t$, and energy label $\alpha$. Here $r$ denotes the distance between the clusters, $s$ is the channel spin, and $\nu$ is a collective index of the relevant quantum numbers. In general, the $^4$He+$^4$He cluster should also be included in the expansion of Eq.~(\ref{eq:rgm_state}). In fact, $^8$Be is unbound and decays into $^4$He+$^4$He nuclei. For technical reasons, we have not included the $^4$He+$^4$He mass partition in the present work (see however our recent progress in the microscopic description of the $^4$He-$^4$He scattering~\cite{Kravvaris:2020cvn}). As we focus here on the low-energy proton capture on $^7$Li where resonances have near-zero $\alpha$-width with a decay to the very narrow $0^+$ ground state, the impact of $^4$He+$^4$He channels is expected to be negligible. The coefficients $c_\lambda^{J^\pi}$ and relative-motion amplitudes $\gamma_{\nu}^{J^\pi}(r)$ are found by solving a two-component, generalized Bloch Schr\"{o}dinger equation derived in detail in Ref.~\cite{physcripnavratil}.
The $\hat{\mathcal{A}}_\nu$ term is the inter-cluster antisymmetrizer:
\begin{align}
\hat{\mathcal{A}}_\nu = \sqrt{\frac{(A-1)!}{A!}}\left(1+\sum_{P\neq id}(-1)^pP\right),
\label{eq:antisymmetrizer}
\end{align}
where the sum runs over all possible permutations of nucleons $P$ (different from the identical one) that can be carried out between the target cluster and projectile, and $p$ is the number of interchanges characterizing them. The resulting NCSMC equations are solved using the coupled-channel R-matrix method on a Lagrange mesh~\cite{R-matrix,Baroni2013}.

The input to the present NCSMC calculations is a microscopic Hamiltonian with the nucleon-nucleon (NN) chiral interaction at next-to-next-to-next-to leading order (N$^{3}$LO) with a cutoff
$\Lambda{=}500$ MeV developed by Entem and Machleidt~\cite{Entem2003}, denoted as
NN-N$^{3}$LO(500). In addition to the two-body interaction, a three-body (3N) interaction at next-to-next-to leading order (N$^{2}$LO) with simultaneous local and nonlocal regularization ~\cite{Navratil2007,Gysbers2019NatPhys,Soma2020} is included. The whole interaction (two- and three-body) will be referred to as NN+3N(lnl). A faster convergence of our NCSMC calculations is obtained by softening the Hamiltonian through the similarity renormalization group (SRG) technique~\cite{Wegner1994,Bogner2007,PhysRevC.77.064003,Jurgenson2009}.
The SRG unitary transformation induces many-body forces that we include up to the three-body level. Four- and higher-body induced terms are small at the $\lambda_{\mathrm{SRG}}=2.0$ fm$^{-1}$ resolution scale used in the present calculations~\cite{PhysRevC.103.035801}. The NCSM and NCSMC basis basis size is characterized by the parameter $N_{\rm max}$ defined as the number of HO excitations above the lowest allowed Pauli configuration. Further, the basis depends on the HO frequency selected to be $\hbar\Omega=20$ MeV for which the ground state energies of the investigated nuclei present a minimum. We note that the microscopic Hamiltonian, the SRG resolution scale, as well as the selected HO frequency are the same as those used for $^{8,9}$Li calculations in Ref.~\cite{PhysRevC.103.035801}. For technical reasons, we are not able to reach basis sizes beyond $N_{\rm max}=9$ in the NCSMC calculations for $^{8}$Be.

\subsection{Radiative Capture}\label{sec:radcap}

\begin{figure}[h]
	\begin{center}
        \includegraphics{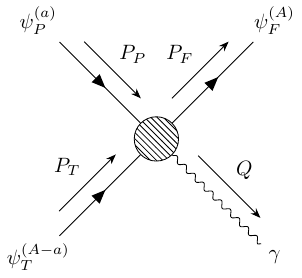}
	\end{center}
	
	\caption{The Feynman diagram and kinematics for radiative capture. $P$ and $Q$ are four-momenta of the nuclei and photon respectively, where e.g. $P_F=(E_F/c,\vec{p}_F)$ and $Q=(\omega/c,\vec{q})$.}
	\label{fig:feyn_cap}
\end{figure}

The process $\ce{^7Li}(p,\gamma)\ce{^8Be}$ is a fusion reaction known as radiative capture.
The Feynman diagram for a general radiative capture nuclear reaction $T(P,\gamma)F$ is presented in
\autoref{fig:feyn_cap}, in which a projectile ($P$) collides with a target ($T$). They fuse, resulting in the composite nucleus ($F$) and the emission of a photon ($\gamma$).

The initial scattering state and final bound state can both be described effectively by the NCSMC formalism. In either case the coefficients of Eq. (\ref{eq:ncsmc}) are found for a given total energy. If this energy is under the break-up threshold (here for $p+\ce{^7Li}$), asymptotics of a bound state are given by Whittaker functions. In contrast, for energies above the threshold, the asymptotics of a scattering state are given by Coulomb functions and a different set of coefficients are found with respect to each possible initial channel $\nu_i$ (describing the cluster decomposition) and partial wave denoted by the orbital, spin and total angular momentum quantum numbers ($\ell_i$, $s_i$ and $J_i$ respectively). The partial waves given in \autoref{tab:p7Li_partialwaves} contribute to the initial state of our calculations for the capture to the ground state.

\begin{table}[tbph]
    \centering
    \begin{tabular}{c|c|c|c|c|c|c|c|c|c|c|}
        $J_i^{\pi_i}$ & \multicolumn{3}{c|}{$1^-$} & \multicolumn{3}{c|}{$1^+$} & \multicolumn{4}{c|}{$2^+$} \\
        \hline
        $s_i$    & 1 & 1 & 2 & 1 & 2 & 2 & 1 & 1 & 2 & 2 \\
        $\ell_i$ & 0 & 2 & 2 & 1 & 1 & 3 & 1 & 3 & 1 & 3 \\
        \hline
    \end{tabular}
    \caption{The included partial waves of the initial $p+\ce{^7Li}$ scattering state for $E1$, $M1$ and $E2$ transitions to the $\ce{^8Be}$ $\left(0^+\right)$ ground state. The spin and parity of the proton $\left(\frac{1}{2}^+\right)$ couple with those of the ground state of $\ce{^7Li}$ $\left(\frac{3}{2}^-\right)$.}
    \label{tab:p7Li_partialwaves}
\end{table}

The transition operator for photon emission \cite{Walecka} is expanded in a linear combination of the transverse electric and magnetic response operators (\ref{eq:T_E}, \ref{eq:T_M}), i.e.
\begin{equation}
    -\vec{e}_\lambda^* \cdot \vec{\mathcal{J}}(q) = \sum_j (-i)^j \sqrt{2\pi} \hat{j} \left[ \mathcal{T}^E_{j-\lambda}(q) +\lambda \mathcal{T}^M_{j-\lambda}(q) \right] \;,\label{eq:eJ_1}
\end{equation}
where $\lambda$ is the photon polarization, $\hat{j}{=}\sqrt{2j+1}$, and $q=|\vec{q}|$ is the photon momentum (for a physical photon $q=\omega/c$ where $\omega$ is the total energy of the transition determined by the difference between the initial kinetic energy and final state binding energy).

The differential cross section for radiative capture can be calculated by evaluating this transition operator between the initial scattering state wavefunctions $\ket{\Psi_{\nu_i}^{(m_T,m_P)}}$ and final bound state wavefunction $\ket{\Psi^{J_f^{\pi_f}M_f}}$, i.e.
\begin{align}
    \frac{\dd\sigma}{\dd\Omega} =& \frac{q}{(2\pi)^2 \hbar v}  \bar{\sum_{m_T m_P}} \sum_{\lambda M_f} \nonumber \\
    &\times \left| \bra{\Psi^{{J_f}^{\pi_f}M_f}} (-\vec{e}_\lambda^* \cdot \vec{\mathcal{J}}(q)) \ket{\Psi_{\nu_i}^{(m_T,m_P)}} \right|^2
    \label{eq:radcap_theory}  \;,
\end{align}
where $v$ is the initial $P$-$T$ relative velocity. The initial state spin projections $m_T$ and $m_P$ are averaged while the final state projections $M_f$ are summed. $J_f$ is the total angular momentum quantum number of the final bound state while $s_P$ and $s_T$ are the spin quantum numbers of the projectile and target respectively. The sum over $\lambda$ runs over the two transverse polarizations of the photon, i.e. $\lambda=\pm 1$.

The wavefunction for the initial scattering state is expanded over a coupled basis and can be written with the normalization and Coulomb phase $\sigma_{\ell_i}$ explicit, i.e.
\begin{align}
    \ket{\Psi_{\nu_i}^{(m_T,m_P)}} =& \frac{\sqrt{4\pi}}{k} \sum_{\ell_i s_i J_i}  i^{\ell_i} \hat{\ell_i} e^{i\sigma_{\ell_i}} \ket{\Psi^{J_i^{\pi_i}}_{\nu_i  s_i \ell_i}} \nonumber \\
    & \times \left( s_T m_T s_P m_P | s_i (m_T+m_P) \right) \nonumber \\
    & \times  \left(s_i (m_T+m_P) \ell_i 0| J_i (m_T+m_P) \right) \;, \label{eq:cluster_state}
\end{align}
where $k= \sqrt{2\mu E_{\rm kin}}/\hbar$ is the relative $P$-$T$ wavenumber, with $\mu$ the reduced mass and $E_{\rm kin}$ the relative kinetic energy. The notation $\hat x$ stands for $\sqrt{2x+1}$. See also Eqs.~(\ref{eq:initial_states}) and (\ref{eq:initial_state_internal}) and the corresponding discussion in Appendix~\ref{sec:radcap_appendix}.

Making use of the low-energy expansion of the electromagnetic operators (\ref{eq:longwave_approx}) and summing over the photon polarizations, the photon production total cross section can be approximated by
\begin{align}
    \sigma =& \frac{8\pi}{vk^2}\frac{1}{\hat{s}_T^2\hat{s}_P^2} \nonumber \\
    &\times\sum_{\kappa j}\frac{\left(\frac{\omega}{\hbar c}\right)^{2j+1}}{[(2j+1)!!]^2}\frac{j+1}{j}\sum_{\ell_i s_i J_i}\left| \mathcal{M}^{\kappa j}_{J_f s_i \ell_i  J_i}\right|^2 \;,
    \label{eq:radcap_longwave_total}
\end{align}
where $j$ is the multipolarity of the electric ($\kappa=\,$E) or magnetic ($\kappa=\,$M) transition operator. 
See Appendix \ref{sec:radcap_appendix} for a detailed derivation, with the resulting Eq. (\ref{eq:radcap_cross}) equivalent to (\ref{eq:radcap_longwave_total}). Comparing this expression to, e.g., Eq.~(1) in Ref.~\cite{Hebborn2022}, the differences are due to alternative conventions in the scattering wave function~(\ref{eq:cluster_state}) normalization.

The matrix elements are defined by:
\begin{align}
    \mathcal{M}^{\kappa j}_{J_f  s_i \ell_i  J_i} &= \bra{\Psi^{J_f \pi_f}}\left|\mathcal{M}^{\kappa}_j\right|\ket{\Psi^{J_i \pi_i}_{ s_i \ell_i}} \label{eq:M_me}  \;,
\end{align}
where we evaluate the $E1$, $E2$ and $M1$ operators, which read
\begin{eqnarray}
    \mathcal{M}^E_{jm} &= e \sum_{i=1}^A \frac{1+\tau_{zi}}{2} r_i^j Y_{jm} (\Omega_i) \label{eq:M_E} \;,\\
    \mathcal{M}^{M}_1 &=\frac{\mu_N}{\hbar c}\sqrt{\frac{3}{4\pi}} \sum_{i=1}^A (g_{\ell i} L_i+g_{si}S_i) \label{eq:M_M}  \;,
\end{eqnarray}
where $e$ is the elementary electric charge, $\mu_N$ is the nuclear magneton, $r_i$ and $\Omega_i$ are the single-nucleon position relative to the center-of-mass (c.m.) of the $A$-nucleon system, $g_{si}$,  $\tau_{iz}$,  $S_i$ and $L_i$ are respectively the gyromagnetic factor, the isospin,   spin and orbital angular momentum (defined with respect to the c.m.) operators of the $i$th nucleon and  $g_{\ell i}$ is $1$ for protons and $0$ for neutrons.

\subsection{Internal \texorpdfstring{$e^+e^-$}{e+e-} Pair Production}\label{sec:pairprod}

\begin{figure}[tbph]
	\begin{center}
        \includegraphics{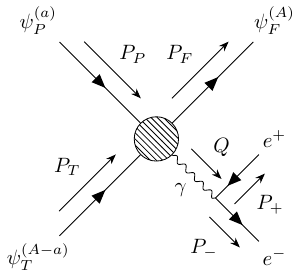}
	\end{center}
	\caption{The Feynman diagram and kinematics for radiative capture with pair
	production. $P_+=(E_+/c,\vec{p}_+)$ and $P_-=(E_-/c,\vec{p}_-)$ are the four-momenta of the positron and electron respectively. The four-momenta of photon is integrated out and hence $Q=(\omega/c,\vec{q})=P_++P_-$ is fixed and represents the total transfer energy $\omega$ and momentum $q$.}
	\label{fig:feyn_cap_pair}
\end{figure}

If the photon emitted in a $\gamma$-decay or radiative capture has
high enough energy ($\omega>2m_{e}c^2$, where $m_e$ is the electron mass) then an electron-positron pair may be produced. Diagrammatically this is shown in 
\autoref{fig:feyn_cap_pair}. The rate is much lower than photon
decays (except in $0^{+}\to0^{+}$ transitions where $\gamma$ decays
are forbidden \cite{Walecka1962,Friar1975}) but recent experiments
have been able to measure the rate, in particular in radiative capture
reactions \cite{Krasznahorkay2016} where additional energy is provided
by the initial kinetic energy of the projectile. The relevant nuclear operators and
kinematics are analogous to electron scattering \cite{deForest1966,Walecka_electron}.

The amplitude for pair production involves the interaction between the charged nuclear current ($\mathcal{J}_\mu$) and leptonic current ($\ell_\mu$).
We evaluate the matrix elements of the charged nuclear current density operators $\mathcal{J}_\mu(q)=\left(\rho(q),\vec{\mathcal{J}}(q)\right)$ through the multipole expansion described in Appendix~\ref{sec:multipole_operators}.
The lepton current is defined with (anti-)leptons modelled via Dirac spinors ($v(P_+)$)$u(P_-)$ and a Dirac gamma matrix $\gamma_\mu$ corresponding to the photon-lepton vertex of Fig. \ref{fig:feyn_cap_pair}, i.e.
\begin{equation}
    \ell_\mu = \bar{u}^{s_-}(P_-)(ie\gamma_\mu)v^{s_+}(P_+) \;.
\end{equation}
Via a sum over the outgoing spinor indices, the lepton tensor $\ell_{\mu\nu}=\sum_{s_+s_-}\ell_\mu \ell_\nu^\dagger$
is simplified into a function of four-momenta which combine into six kinematic prefactors of different combinations of nuclear transition matrix elements (after summation over initial and final states and integration over the intermediate photon momentum). 
See Appendix~\ref{sec:pairprod_radcap_appendix} for details.

The differential cross section for pair production in radiative capture, describing the probability distribution of pair energies and angles per unit input flux, can be simplified into the expression:
\begin{align}
    \frac{\dd^5\sigma}{\dd E_+ \dd\Omega_+ \dd\Omega_-} &= \frac{1}{32 \pi^5 v}\frac{p_+p_-}{Q^4} \sum_{n=1}^{6} v_n R_n\;, \label{eq:dsigma_theory}
\end{align}
equivalent to Eq. (\ref{eq:full_diff_sigma}) in Appendix~\ref{sec:pairprod_radcap_appendix}, where $E_+$ is the energy of the positron (the energy of the electron $E_-$ is fixed by energy conservation). 
$\Omega_+$ and $\Omega_-$ are the solid angles into which the positron and electron are emitted.
The electron and positron momenta, $p_+$ and $p_-$, are determined from $E_+$ by their dispersion relations.
Each term in the sum is a combination of a kinematic factor ($v_n$) and a product of nuclear transition matrix elements ($R_n$), consistent with the notation of \cite{Viviani2022}.
In this expression we neglect the recoil of the final nucleus and present the kinematic factors in natural units (suppressing factors of $\hbar$ and $c$).
Note that the term $R_n$ is proportional to $e^4= 16\pi^2\alpha^2$ whereas the radiative capture squared amplitude is proportional to $e^2=4\pi\alpha$.

\subsection{Hypothetical Intermediate Particles\label{subsec:Hypothetical-X-Op}}

In the low-energy limit, the cross section for emission of a hypothetical boson in radiative capture is calculated by inserting a new operator $\mathcal{O}^{(X)}$ into the transition matrix element of Eq. (\ref{eq:radcap_longwave_total}) (using the same notation as Eq. (\ref{eq:M_me})), i.e.
\begin{equation}
        \sigma = \frac{8\pi q_X}{\hbar vk^2}\frac{1}{\hat{s}_P^2\hat{s}_T^2}\sum_{\ell sJ}\left| \mathcal{O}^{(X)}_{J_f s \ell J}(q_X)\right|^2
\end{equation}
where $q_{X}=\sqrt{\omega^{2}-m_{X}^{2}c^4}/c$ is the momentum carried by the new particle and $\omega$ is the total transition energy.
We consider only the leading
multipole $j=1$.
Any possible interference between $\gamma$ and $X$ is not considered in the present calculations.

For a vector particle, we set the operator proportional to the electric dipole operator, $E1$
\cite{Feng2020}, i.e.
\begin{align}
	\mathcal{O}^{(V)}(q_X)= & \epsilon_{V}\frac{q_{X}}{e\hbar}\mathcal{M}_{1}^{E}\;.
 \label{eq:V}
\end{align}

For a pseudo-scalar (axion-like) particle, the operator \cite{Donnelly1978} reduces to the nuclear spin in the long-wavelength limit, i.e.
\begin{align}
	\mathcal{O}^{(P)}(q_X)= & \epsilon_{P}\frac{q_{X}}{\hbar}\sum_{i=1}^A S_i\;.
 \label{eq:P}
\end{align}

The operator for an axial vector particle is also proportional to the nuclear spin \cite{Kozaczuk2017}, i.e.
\begin{align}
	\mathcal{O}^{(A)}(q_X)= & \epsilon_{A}\sqrt{2+\frac{\omega^{2}}{m_{X}^2c^4}}\sum_{i=1}^A S_i\;.
 \label{eq:A}
\end{align}

In principle, the couplings $\epsilon_{X}$ would be fitted to match
experiments.

\section{Results}\label{results}
\subsection{NCSM calculations for \texorpdfstring{$^7$Li}{7Li} and \texorpdfstring{$^{7,8}$Be}{7,8Be}}\label{NCSM_results}

The ansatz of the present NCSMC calculations requires the NCSM eigenstates and eigenenergies of $^{7}$Li and $^{7,8}$Be. For the expansion in Eq.~\eqref{eq:ncsmc}, we used the $^{7}$Li and $^{7}$Be $3/2^-$ ground states and the first four excited  states, $1/2^-, 7/2^-, 5/2^-, 5/2^-$, and the lowest 15 (15) positive (negative) parity eigenstates for $^{8}$Be with $J$ ranging from $0$ to $6$. We have performed the NCSM calculations up to $N_{\rm max}{=}10$. However, for technical reasons only up to $N_{\rm max}{=}9$ eigenstates were used in the NCSMC calculations. 

\begin{figure*}[tbph]
    \centering
    \begin{subfigure}{0.5\textwidth}
        \centering
        \includegraphics[width=0.95\linewidth]{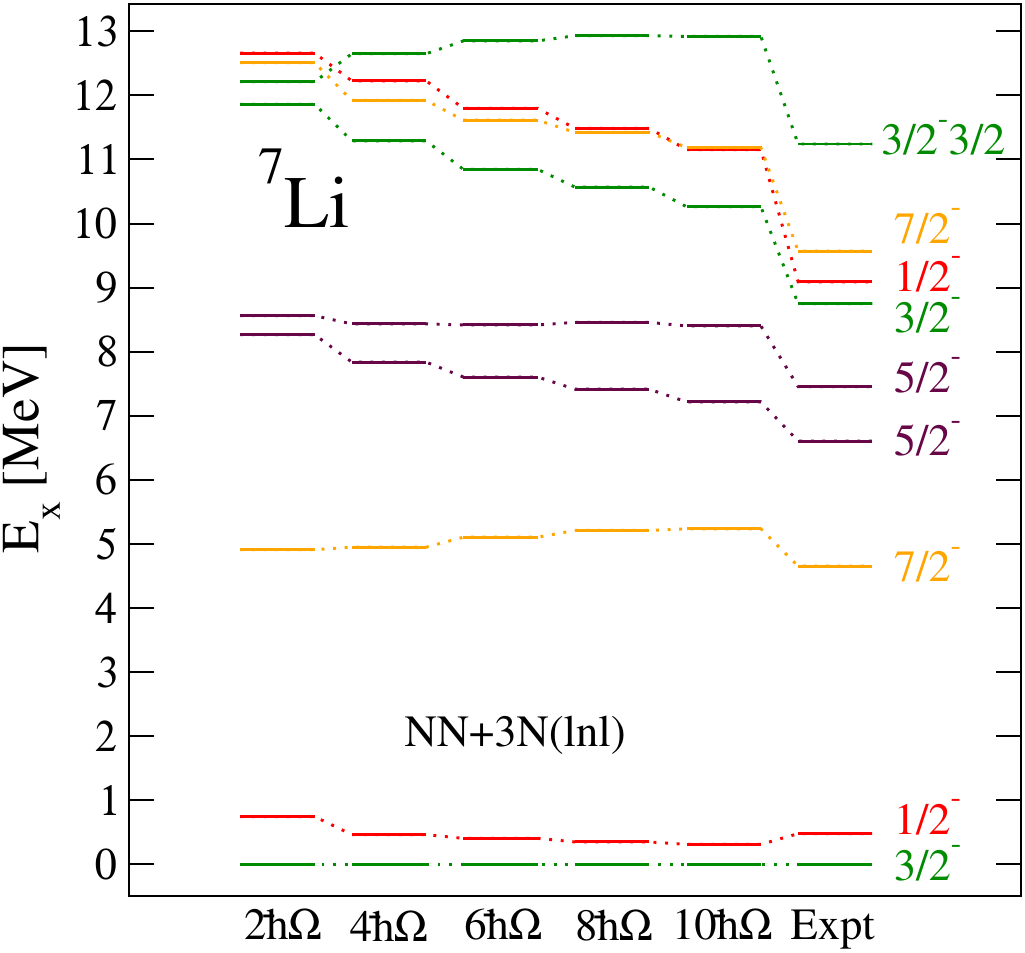}
               \caption{}
        \label{fig:7Li_NCSM}
    \end{subfigure}%
    \begin{subfigure}{0.5\textwidth}
        \centering
        \includegraphics[width=0.95\linewidth]{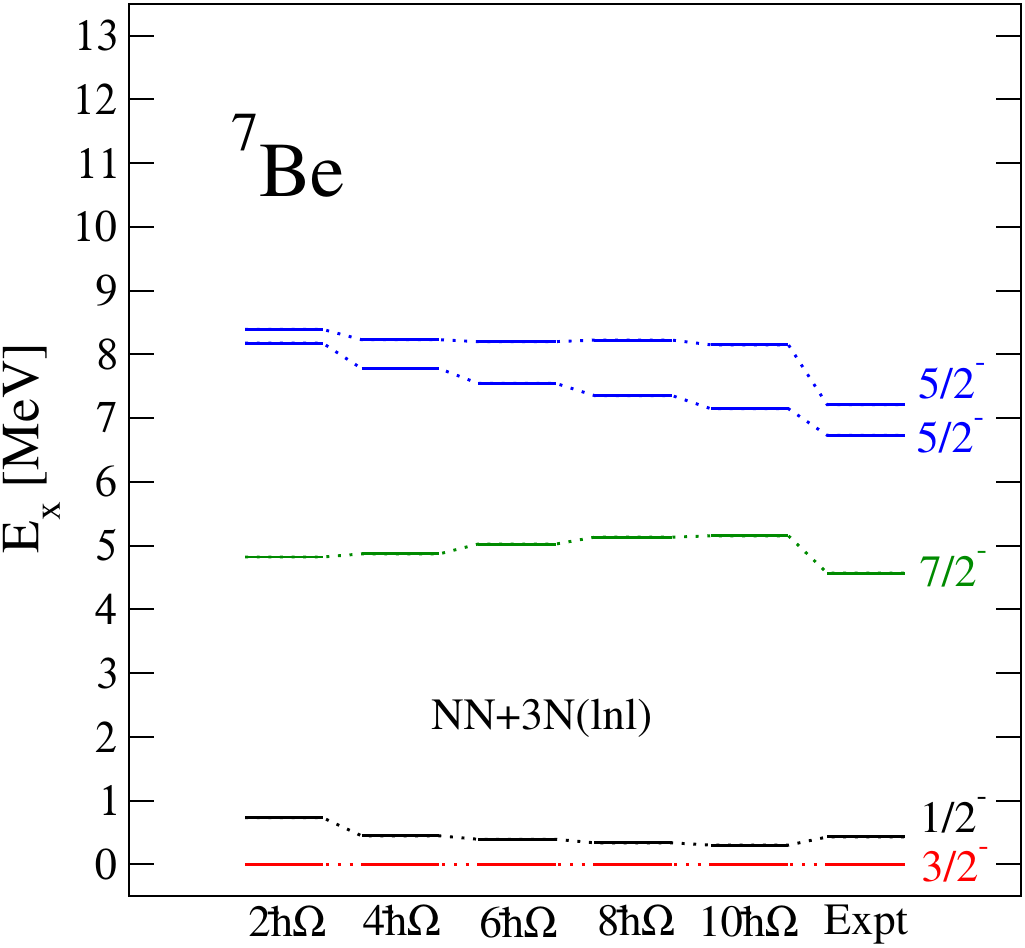}
              \caption{}
        \label{fig:7Be_NCSM}
    \end{subfigure}
    
    \begin{subfigure}{0.5\textwidth}
        \centering
        \includegraphics[width=0.95\linewidth]{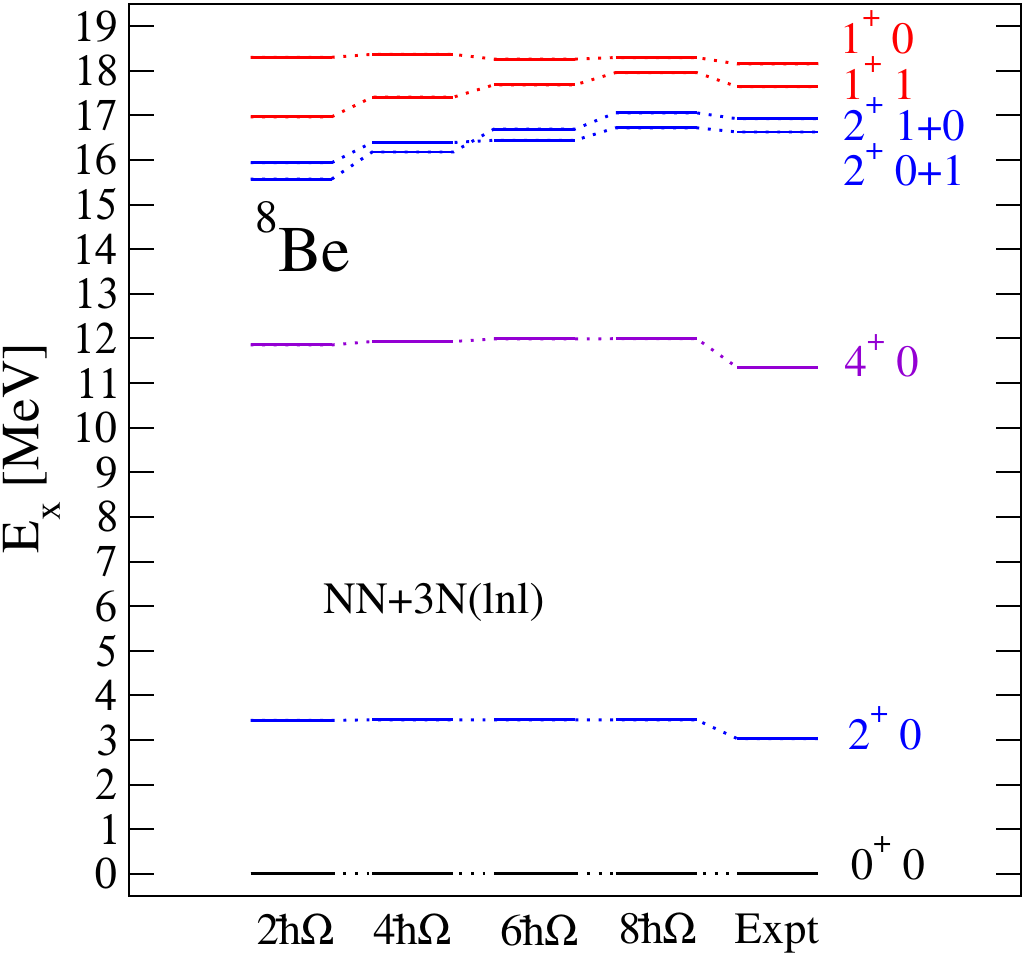}
               \caption{}
        \label{fig:8Be_NCSM}
    \end{subfigure}%
    \begin{subfigure}{0.5\textwidth}
        \centering
        \includegraphics[width=\linewidth]{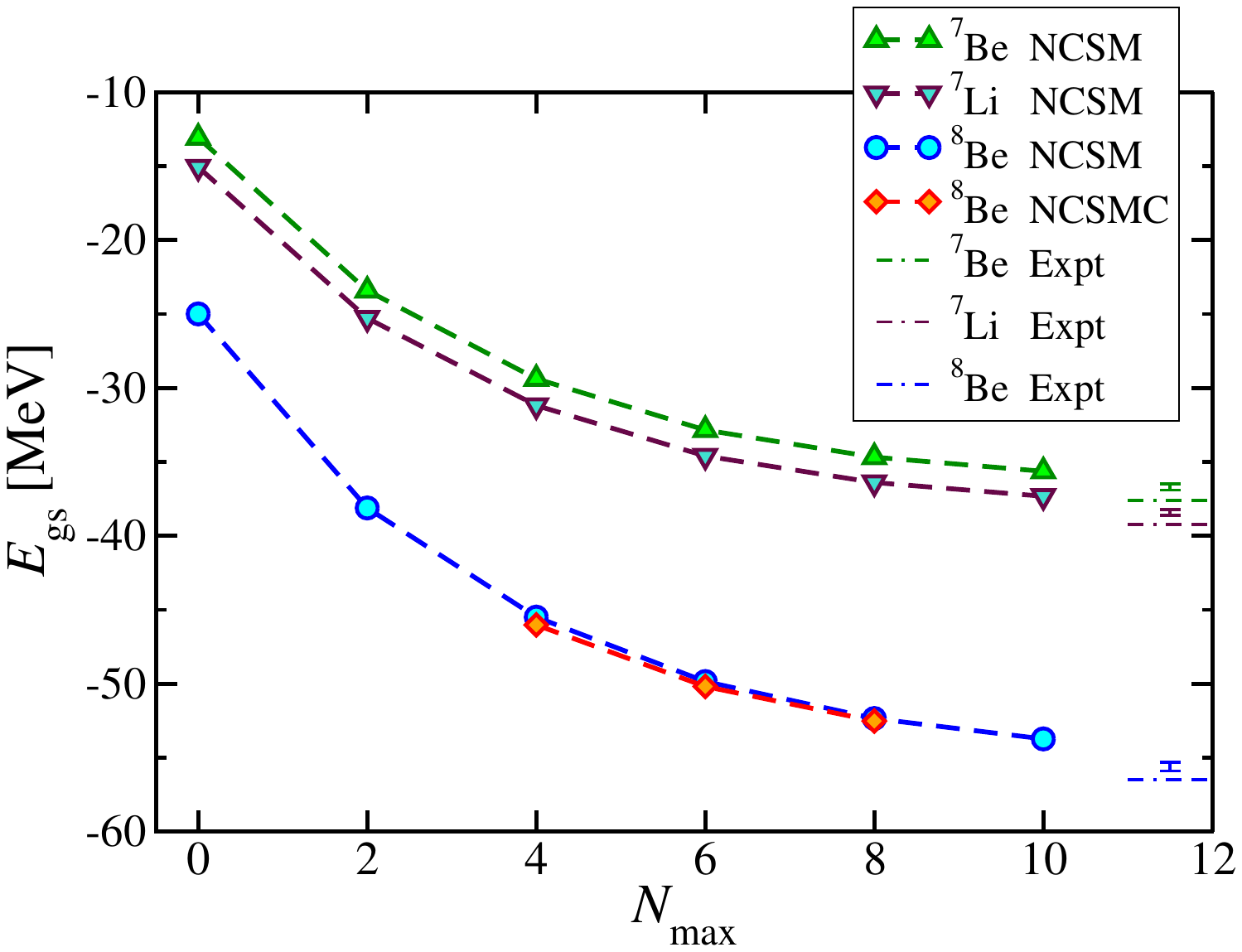}
        \caption{}        \label{fig:gs_convergence}
    \end{subfigure}

	\caption{Excitation energies from the NCSM for (a)$\ce{^{7}Li}$, (b) $\ce{^{7}Be}$
		and (c) $\ce{^{8}Be}$. The lowest states most relevant for input to the NCSMC are in the correct order.
		Only the lowest 5 states are shown for $\ce{^{7}Be}$. (d) The convergence of the NCSM ground state energies of $\ce{^{7}Li}$, $\ce{^{7}Be}$
		and $\ce{^{8}Be}$ with the model-space parameter $N_{\rm max}$. Extrapolations to infinite $N_{\rm max}$ with their uncertainties are presented on the right. For $\ce{^{8}Be}$, NCSMC results (diamonds) are also shown.  The NN+3N(lnl) interaction~\cite{Soma2020} was used. Experimental data are from Ref.~\cite{Tilley2004}. See the text for further details.}
	\label{fig:NCSM_A7_A8_gs}
 \end{figure*}

The dependence of the low-lying excitation energies of $^7$Li, $^7$Be, and $^8$Be on the basis size is shown in Figs.~\ref{fig:7Li_NCSM}, \ref{fig:7Be_NCSM}, and \ref{fig:8Be_NCSM}, respectively. For $^7$Li and $^8$Be, these results were already reported in Ref.~\cite{Soma2020}. For $^7$Be, we only show the four lowest excited states employed in the present NCSMC investigation. For $^8$Be, we show only the four excited states experimentally below the $p{+}^7$Li threshold and the two $1^+$ states above the threshold that are key to the present study for clarity of the figure. It should be noted, however, that we still include eight higher positive parity states not shown in the figure as input for the NCSMC. Overall, the agreement with experiment is satisfactory as is the $N_{\rm max}$ convergence. We also note the correct order of the calculated states compared to experiment. 

 The ground-state energy dependence on the basis size for all three isotopes is presented in Fig.~\ref{fig:gs_convergence}. The extrapolated NSCM $^7$Li, $^7$Be, $^8$Be ground state energies to the infinite basis size are $-38.4(2)$ MeV, $-36.7(2)$ MeV, $-55.6(3)$ MeV, respectively. Comparing to the experimental values of $-39.245$ MeV, $-37.60$ MeV, $-56.50$ MeV, respectively, the computed states are underbound by about 2 percent. The theoretical uncertainty is due to the extrapolation to the infinite basis size performed using the exponential function $E (N_{\rm max}) = E_{\infty} + a e^{- b N_{\rm max}}$ and varying the number of $N_{\rm max}$ points.

\subsection{NCSMC calculations for \texorpdfstring{$^{8}$Be}{8Be}}\label{NCSMC_results}

We performed NCSMC calculations for $^8$Be in $N_{\rm max}{=}4,6,8$ basis spaces. The $^8$Be NCSM negative-parity states entering the expansion~(\ref{eq:ncsmc}) were obtained in $N_{\rm max}{+}1$ spaces, i.e., up to $N_{\rm max}{=}9$. In the following, we use, e.g., $N_{\rm max}{=}8$ and $N_{\rm max}{=}9$ interchangeably to refer to the same set of NCSMC calculations as both spaces contribute simultaneously in the cross section calculations. The NCSMC ground-state energies are shown in Fig.~\ref{fig:gs_convergence} and the separation energies with respect to the $^7$Li${+}p$ threshold as well as the excitation energies for the lowest five states obtained in the $N_{\rm max}{=}8$ space are given in Table~\ref{tab:8Be}. Comparing to the input NCSM calculations, we can see a clear improvement. The lowest four calculated states are below the $^7$Li${+}p$ threshold as in experiment, only the third $2^+$ state is slightly above the threshold. Energies of the $0^+_1$, $2^+_1$, and $4^+_1$ states are overestimated by about 1 MeV, which can be attributed to the neglect of the $^4{\rm He}{+}^4{\rm He}$ mass partition in the present calculations. Because of that, we describe the states below the $^7$Li${+}p$ threshold as bound with a zero width in contrast to the experiment in particular for the $2^+_1$ and $4^+_1$ states. As stated before, the $0^+$ ground state of $^8$Be is experimentally very narrow. The approximation we are using is satisfactory for its description. 

\begin{table}[tbph]
	\centering
		\begin{tabular}{ccccccc}
			& \multicolumn{3}{c}{Energy {[}MeV{]}} & \multicolumn{3}{c}{Excitation Energy {[}MeV{]}}\tabularnewline
			\hline 
			& NCSM & NCSMC & Expt. & NCSM & NCSMC & Expt.\tabularnewline
			\hline 
			\hline 
			$0^{+}$ & -15.96 & -16.13 & -17.25 & 0.00 & 0.00 & 0.00\tabularnewline
			$2^{+}$ & -12.51 & -12.72 & -14.23 & 3.45 & 3.41 & 3.03\tabularnewline
			$4^{+}$ & -3.97 & -4.31 & -5.91 & 11.99 & 11.82 & 11.35\tabularnewline
			$2^{+}$ & +0.76 & -0.10 & -0.63 & 16.72 & 16.03 & 16.63\tabularnewline
			$2^{+}$ & +1.09 & +0.31 & -0.33 & 17.05 & 16.44 & 16.92\tabularnewline
		\end{tabular}
	\caption{NCSM and NCSMC energies of the lowest five states of $\ce{^{8}Be}$ with respect to the $\ce{^{7}Li}+p$ threshold (left) and their excitation energies (right) compared to experiment. The NN+3N(lnl) interaction~\cite{Soma2020} in the $N_{\rm max}{=}8$ space was used. Experimental data are from Ref.~\cite{Tilley2004}. See the text for further details.}
	\label{tab:8Be}
\end{table}

\begin{table}[tbph]
\begin{tabular}{ccccccc}
          & \multicolumn{2}{c}{NCSMC}          & \multicolumn{2}{c}{NCSMC-pheno} & \multicolumn{2}{c}{Expt.}  \\
\hline
$J^\pi_n$ & $E_{\rm res}$  & $\Gamma$  & $E_{\rm res}$  & $\Gamma$  & $E_{\rm res}$  & $\Gamma$  \\
\hline
\hline
$1^+_1$   & 0.734         & 0.0899             & 0.390    & 0.0126  & 0.385(1) & 0.0107(5)\\
$1^+_2$   & 1.098         & 0.1332             & 0.905    & 0.1083  & 0.895(4) & 0.138(6) \\
$1^+_3$   & 2.788         & 0.7889             & 2.501    & 0.468   & N/A & N/A  \\
$1^+_4$   & 2.868         & 3.0094             & 2.933    & 3.047   & N/A & N/A \\
$3^+_1$   & 2.646         & 0.7941             & 2.376    & 0.5669  & 1.815(3) & 0.270(20) \\
$3^+_2$   & 2.868         & 0.3894             & 2.546    & 0.2579  & 1.980(10)& 0.227(16)     
\end{tabular}
\caption{NCSMC and NCSMC-pheno resonance centroids ($E_{\rm res}$) and widths ($\Gamma$), in MeV, of $1^+$ and $3^+$ states compared to experiment~\cite{Tilley2004}. See text for details.}
\label{tab:1p3pres}
\end{table}

\begin{figure}[tbph]
	\centering
\includegraphics[width=\linewidth]{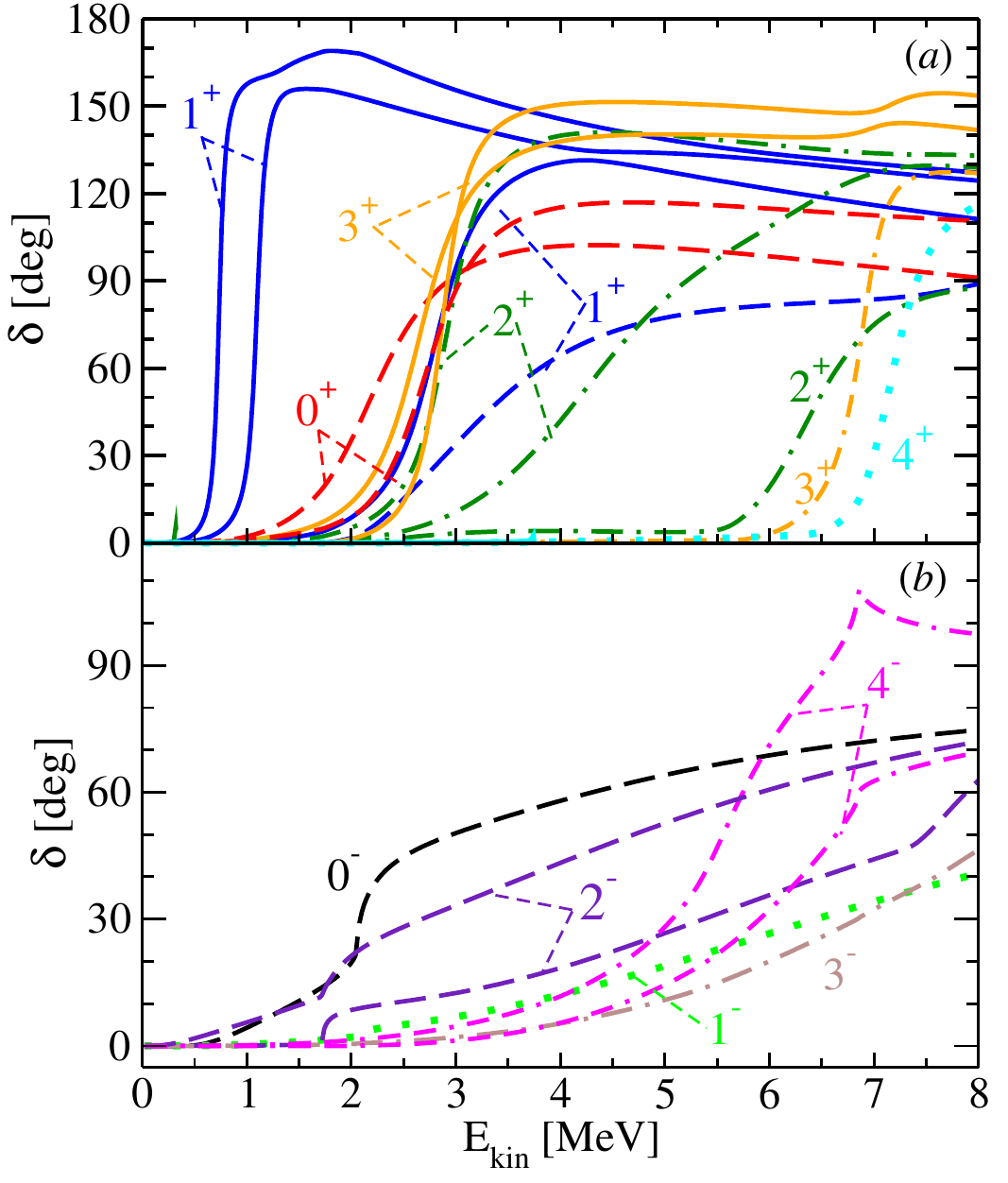}
	\caption{$\ce{^{7}Li}+p$ eigenphase shifts for even (a) and odd (b) parity channels obtained within NCSMC. The NN+3N(lnl) interaction in the $N_{\rm max}{=}9$ space was used. $E_{\rm kin}$ is the kinetic energy of the $\ce{^{7}Li}{+}p$ in the center of mass frame.}
	\label{fig:eigenphases}
\end{figure}
\autoref{fig:eigenphases} shows the eigenphase shifts for scattering with $\ce{^{7}Li}{+}p$ and $\ce{^{7}Be}{+}n$ channels. Calculations for $N_{\rm max}{=}9$ are shown. The eigenphase shifts $\delta$ are calculated from the eigenvalues $e^{2i\delta}$ of the S-matrix. The S-matrix is computed as a function of the energy in the center of mass, shown here up to 8 MeV. The eigenphase shift corresponding to the third $2^{+}$ state, which is below the $\ce{^{7}Li}{+}p$ threshold in experiment (see \autoref{tab:8Be}), shows up as an extremely narrow resonance and can be seen in the bottom-left corner of panel (a). The next two resonances are $1^{+}$ states which correspond to the spin aligned and anti-aligned interactions between the proton and the $\tfrac{3}{2}^{-}$ ground state of $\ce{^{7}Li}$.
These $1^{+}$ resonances are of particular interest for the present investigation as discussed in the Introduction. They correspond to the experimental 17.64 MeV and 18.15 MeV $1^+$ excited states in $^8$Be. We note that the $^7$Li${+}p$ threshold is experimentally at 17.255 MeV, i.e., those two resonances appear at 0.385 MeV and 0.895 MeV above the threshold, respectively. 

Many more resonances are predicted past the threshold of $\ce{^{7}Be}{+}n$ which is higher than $\ce{^{7}Li}{+}p$. We reproduce the $3^{+}$ pair in particular, which is well established experimentally and find several resonances additional to the TUNL evaluation~\cite{Tilley2004}. For example, we find a $0^+$ resonance below the $3^+$ pair and a pronounced $0^-$ resonance near the $\ce{^{7}Be}(1/2^-)+n$ threshold. It is more dominant in our calculation than the $2^-$ resonance established experimentally~\cite{Tilley2004}. The cusps seen in the negative-parity eigen-phaseshifts in Fig~\ref{fig:eigenphases} (b) appear at the thresholds: $2^-$ at $\ce{^{7}Be}(3/2^-)+n$, $0^-$ at $\ce{^{7}Be}(1/2^-)+n$, $4^-$ at $\ce{^{7}Be}(7/2^-)+n$.

The centroid energies with respect to the $^7$Li${+}p$ threshold and widths of the $1^+$ and $3^+$ resonances obtained in the $N_{\rm max}{=}9$ space are presented in Table~\ref{tab:1p3pres}. NCSMC resonance characteristics are extracted from the eigenphase shifts with the centroid position $E_{\rm res}$ given by the inflection point and the width calculated as (see, e.g., Ref.~\cite{ThompsonNunes}) $\Gamma = \frac{2}{\frac{\dd \delta}{\dd E}\vert_{E_{\rm res}}}$ with the $\delta$ in radians. The NCSMC results presented in the second and third column overestimate somewhat the energy and the width of the lowest $1^+$ state. The second $1^+$ state is obtained close to experiment. Similarly, the the $3^+$ resonance positions and widths are overestimated with the second $3^+$ state described better compared to experiment. For completeness, we also show the NCSMC predictions for the third and fourth $1^+$ states that do not have experimental counterparts at present. The fourth state is particularly broad. We do not show the centroids and widths of the $0^+$ and other higher-lying resonances as they will likely be impacted by the $^4$He-$^4$He mass partition not included in the present calculations.

\begin{figure}[tbph]
    \centering
        \includegraphics[width=\linewidth]{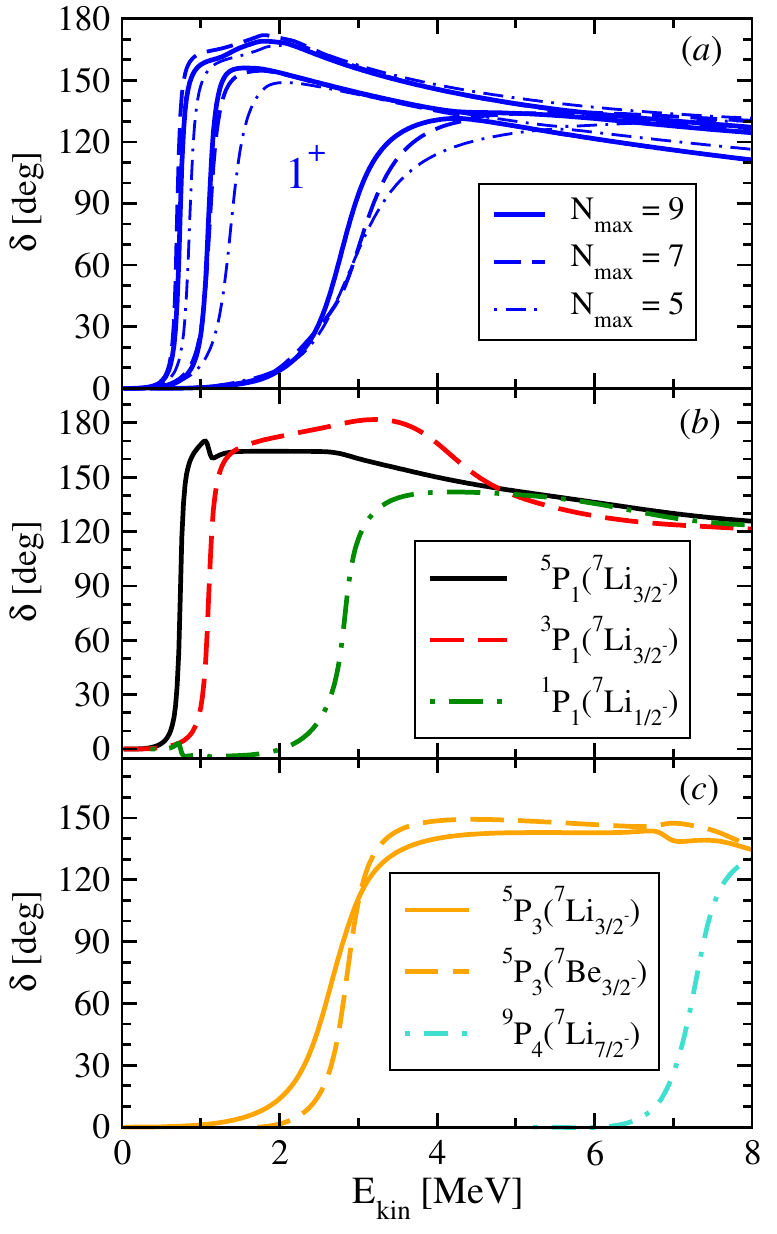}
	\caption{
     (a) Dependence of the $\ce{^{7}Li}+p$ eigenphase shifts in the $1^+$ channel on the NCSMC basis size characterized by $N_{\rm max}$. (b) Selected diagonal $\ce{^{7}Li}+p$ $P$-wave phase shifts in the $1^+$ channel at $N_{max}=9$. (c) Selected diagonal $P$-wave $\ce{^{7}Li}{+}p$ and $\ce{^{7}Be}{+}n$ phase shifts for $J^\pi{=}3^+$ and $4^+$ channels at $N_{\rm max}{=}9$. 
    The NCSMC calculated $\ce{^{7}Be}{+}n$ threshold is at 1.72 MeV compared to the experimental 1.64 MeV. ${\rm E}_{\rm kin}$ is the kinetic energy of the $\ce{^{7}Li}{+}p$ in the center of mass frame.} 
	\label{fig:1plus_phaseshifts}
\end{figure}
\begin{figure}[tbph]
    \centering
    \includegraphics[width=\linewidth]{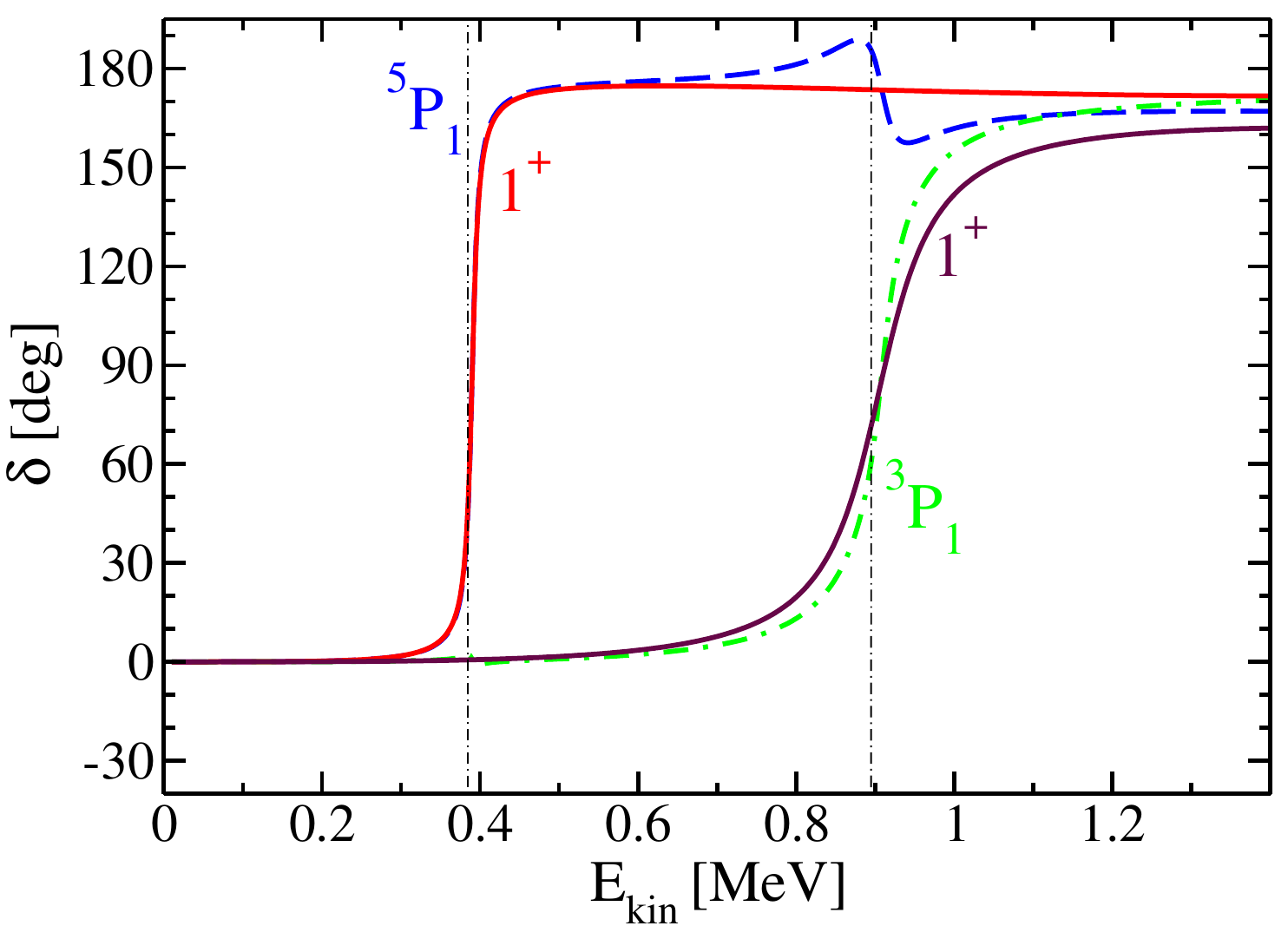}
	\caption{Eigenphase shifts and diagonal phase shifts obtained within the NCSMC-pheno approach at $N_{max}=9$. The vertical dashed-dotted lines mark the experimental resonance positions. See the text for further details.}
    \label{fig:1plus_diagphaseshifts}
\end{figure}

The convergence of the $1^+$ eigenphase shift with $N_{\rm max}$ is shown in Fig. \ref{fig:1plus_phaseshifts} (a). The lowest two resonances show a good stability with respect to the basis size change, i.e., there is little difference for $N_{\rm max}{=}9$ and 7. The structure of the resonances is revealed by examining the diagonal phase shifts shown in Fig. \ref{fig:1plus_phaseshifts} (b). While all three $1^+$ resonances are of $P$-wave character, the lowest two are built on the $3/2^-$ ground state of $^7$Li with the first one with the channel spin $s{=}2$ (parallel spins of $p$ and $^7$Li) and the second with $s{=}1$ (antiparallel spins), see also the zoomed-in~\autoref{fig:1plus_diagphaseshifts}. The third $1^+$ resonance is built on the $^7$Li $1/2^-$ exited state. We can also extract the isospin character of the resonances with the first being predominantly $T{=}1$ and the second $T{=}0$.

The diagonal phase shifts of the two lowest $3^{+}$ resonances and of the lowest $4^+$ resonance, all of $P$-wave character, are shown in Fig.~\ref{fig:1plus_phaseshifts} (c). Interestingly, the first $3^+$ resonance is built on the ground state of $\ce{^{7}Li}{+}p$ while the second one on the ground state of $\ce{^{7}Be}{+}n$. The $4^+$ resonance is built on the $7/2^-$ excited state of $^7$Li. All three resonances have the target and projectile spins aligned.

\subsection{Phenomenologically adjusted NCSMC}\label{sec:pheno}

Before proceeding with the calculation of the capture cross sections, the NCSMC results were phenomenologically adjusted to reproduce experimental thresholds and positions of the two $1^+$ resonances (i.e., their energies, not widths) in an approach known as NCSMC-pheno ~\cite{DohetEraly2016,Calci2016}. 
This step is necessary to obtain a quantitative evaluation of the capture cross section. The resulting evaluation embodies an advanced microscopic understanding of the underlying nuclear structure and reaction mechanism obtained from a chiral NN+3N Hamiltonian, but is no-longer a purely theoretical prediction. 

We performed the NCSMC-pheno calculations for the $N_{\rm max}{=}8$ model space. The phenomenological modifications are rather small and were accomplished first by adjusting the NCSM calculated $\ce{^{7}Be}{+}n$ threshold 1.7189 MeV with respect to the $\ce{^{7}Li}+p$ obtained at $N_{\rm max}{=}8$ to the experimental value of 1.6446 MeV. Second, we adjust the excitation energies of $^7$Li and $^7$Be to the experimental values, again a rather small modification (with the largest shifts of ${\approx} 1$~MeV for the second $5/2^-$ states) as the NCSM reproduces these energies reasonably well as seen in Fig.~\ref{fig:7Li_NCSM} and~\ref{fig:7Be_NCSM}. All these energies serve as input to the NCSMC. Third, we fit the $^8$Be NCSM input energies to reproduce the experimental $^8$Be energies in the NCSMC calculations. In particular, we fit the two $1^+$ resonances of $^8$Be. Further, the $2^+_1$, $2^+_2$ and $2^+_3$ states were brought closer although not exactly to their experimental positions. In particular, the $2^+_3$ state became bound. No other $J^\pi$ channels were adjusted. 

The impact on the two $1^+$ resonances of interest can be judged by examining the NCSMC and NCSMC-pheno results presented in Table~\ref{tab:1p3pres} and by comparing Figs.~\ref{fig:1plus_phaseshifts} (b) and ~\ref{fig:1plus_diagphaseshifts}. As seen in Table~\ref{tab:1p3pres}, the widths of the two lowest $1^+$ states predicted by the NCSMC-pheno are in a good agreement with experiment. The $3^+$ resonance centroids and widths are also closer to experiment compared to the original NCSMC calculations. As discussed above, the $3^+$ resonance energies have not been fitted to experiment unlike the $1^+$ resonance energies and the changes seen in Table~\ref{tab:1p3pres}, all in the direction towards experiment, are due to the adjustments of the $^7$Li and $^7$Be energies. 

\subsection{\texorpdfstring{$p+\ce{^{7}Li}$}{p+7Li} Scattering}

The NCSMC formalism allows to describe simultaneously the structure of $^8$Be as well as nuclear reactions of sub-clusters. The simplest process relevant to the present study is the elastic proton scattering on $^7$Li. The differential cross section can be obtained from
the NCSMC-computed multi-channel S-matrix. 

\begin{figure*}[tbph]
    \centering
    \begin{subfigure}{0.5\textwidth}
        \centering
        \includegraphics[width=1.0\textwidth]{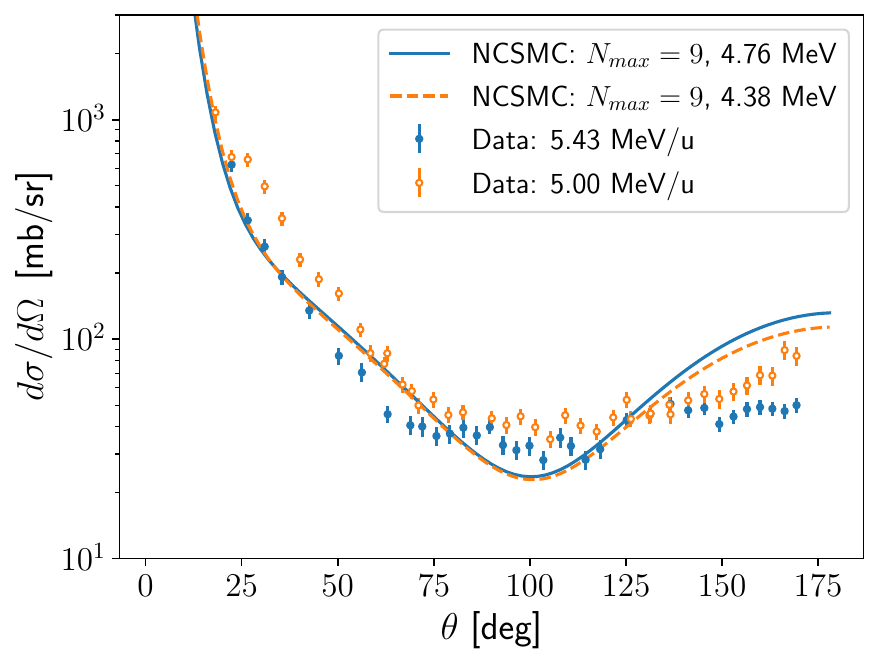}
    \end{subfigure}%
    \begin{subfigure}{0.5\textwidth}
        \centering
        \includegraphics[width=1.0\textwidth]{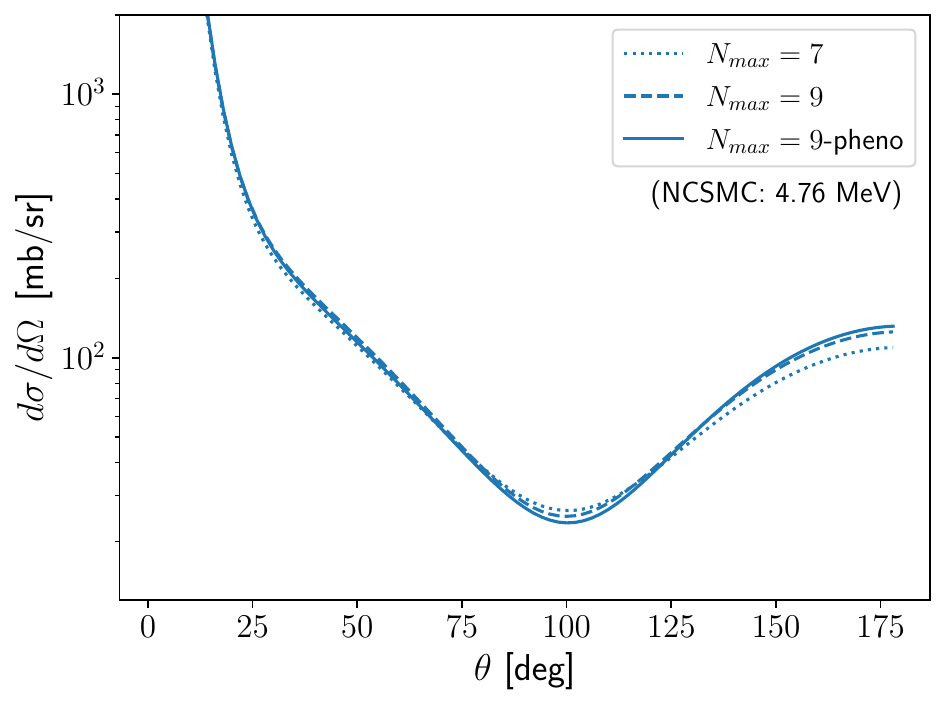}
    \end{subfigure}
	\caption{The differential cross section for $\ce{^7Li}(p,p)\ce{^7Li}$ elastic scattering. $\theta$ is the angle between the incoming proton and outgoing proton in the center of mass frame. (a) A comparison between inverse scattering data at $5$ and $5.44$ MeV/u and NCSMC calculations at the corresponding center of mass energies: $E=4.38$ and $4.76$ MeV. Data from \cite{Pakou2016,Pakou2017}. (b) The convergence of NCSMC calculations at $E=4.76$ MeV with the input parameter $N_{\rm max}$. The effect of the phenomenological correction described in Section~\ref{sec:pheno} is also shown.}
	\label{fig:p7Li_scattering}	
\end{figure*}
As an example, we show the differential cross section for $\ce{^{7}Li}(p,p)\ce{^{7}Li}$ elastic scattering in Figure \ref{fig:p7Li_scattering}. The calculation is compared to data from inverse kinematics experiments i.e., $p(\ce{^{7}Li},\ce{^{7}Li})p$. The lab energy is $5$ and $5.44$ MeV per nucleon (i.e. $\ce{^{7}Li}$ with a beam energy of $35$ and $38.1$~MeV)~\cite{Pakou2017,Pakou2016}. These energies are beyond the point at which phenomenological shifts of the resonances are applied as described in Section~\ref{sec:pheno}. Our calculations predict a very large number of resonances and their individual contributions to the cross sections have not yet been fully analyzed. Up to about 125$^\circ$, the calculations reasonably reproduce the data. For larger angles, the cross section is overestimated, which is likely caused by incorrect positions of resonances in the scattering energy range. In panel (b), the dependence of the differential cross-section on $N_{\rm max}$ and the pheno correction is shown. Clearly, the $N_{\rm max}$ dependence is small and the impact of the phenomenological corrections minimal.

The cross sections for inelastic scattering and charge-exchange reactions, $\ce{^{7}Li}(p,n)\ce{^{7}Be}$ and $\ce{^{7}Be}(n,p)\ce{^{7}Li}$, are also obtained in our formalism. Results will be presented and compared to data in a future publication.

\subsection{Radiative Capture}\label{radcapture}

Using Eq.~(\ref{eq:radcap_longwave_total}), we calculate the radiative
capture cross section between the initial cluster scattering state
and the final $\ce{^{8}Be}$ $0^{+}$ ground state or the $2^{+}_1$ excited state both treated as bound, considering $E1$, $M1$ and $E2$ multipolarities.

\begin{figure}[tbph]
    \centering
    \includegraphics[width=\linewidth]{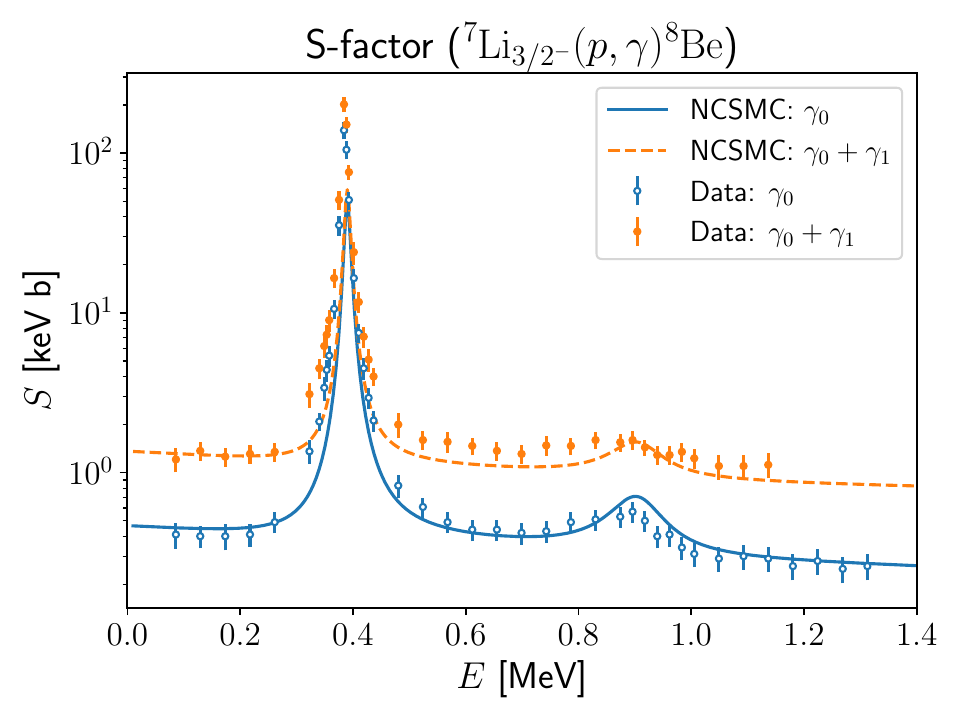}
    \caption{
        The $S$ factor for $\ce{^{7}Li}(p,\gamma)\ce{^{8}Be}$ calculated at $N_{\rm max}=9$ within NCSMC-pheno.  $\gamma_{0}$ indicates proton capture resulting in the ground state of $\ce{^{8}Be}$ while $\gamma_{1}$ is the contribution to the capture to the first excited state ($2^{+}$) of $\ce{^{8}Be}$. Data points are taken from \cite{Zahnow1995}.}
    \label{fig:capture}
\end{figure}
In \autoref{fig:capture}, we compare the NCSMC-pheno predictions of astrophysical $S$ factor proportional to the integrated cross section to the experimental data of Ref. \cite{Zahnow1995}. The $S$ factor is defined by
\begin{equation}\label{Sfact}
    S(E)=\sigma(E)\; E \; {\rm exp}(2\pi\eta) \; ,
\end{equation}
with $\eta{=}Z_P Z_T \; e^2/\hslash v$ the Sommerfeld parameter describing the $S$-wave barrier penetration and $v{=}\sqrt{2E/\mu}$ the relative target-projectile velocity.
After phenomenological adjustment the calculations match the data
very well. The adjustment shifts the position of the peaks but
the size of the background ($E1$) transition strength is already fairly well predicted in the original calculation (not shown in the figure), indicating that the \emph{ab initio} wave functions are quite realistic.

For much of the energy range the dominant term in the integrated cross-section
is the $E1$ transition strength, corresponding to transitions from
the $1^{-}$ partial wave, in the case of transitions to the ground state. However within the $1^{+}$ resonances, the $M1$ transition
strength is greatly enhanced. This is in particular true for the $1^+_1$ resonance which is dominated by isospin $T{=}1$ and allows an isovector decay to the $T{=}0$ ground state and the $2^+_1$ state of $^8$Be. The second $1^+$ resonance is dominated by the isospin $T{=}0$ and thus the $M1$ decays are isospin suppressed. Still, we find some $M1$ enhancement above the $E1$ background. For the spin-spatial structure of these two resonances, see Figures~\ref{fig:1plus_phaseshifts} (b) and ~\ref{fig:1plus_diagphaseshifts} and the discussion in Sect.~\ref{NCSMC_results}. The $E1$ and $M1$ contributions to the integrated $\gamma$-capture cross section are shown in Figs.~\ref{fig:pairprod_2022} (b) and \ref{fig:capture_X}. The $E2$ transition strength is negligible at the order of 1\% (shown in Fig.~\ref{fig:pairprod_2022} (b)).

\subsection{Pair Production}\label{pairprod}

Applying the formalism discussed in Sect.~\ref{sec:pairprod} and in the Appendix~\ref{sec:pairprod_radcap_appendix}, we calculate the internal $e^+ e^-$ pair conversion differential cross section (i.e., Eq.~(\ref{eq:partial_sigma})). The cross section is dominated by the $E1$ and $M1$ contributions. Using the relation (\ref{eq:Siegert}) and the approximations of (\ref{eq:longwave_approx}) we use electric multipoles in the place of Coulomb and longitudinal current operators. The $E2$ contribution is suppressed by several orders of magnitude. 

\begin{figure*} 
	\centering
    \includegraphics[width=1.0\textwidth]{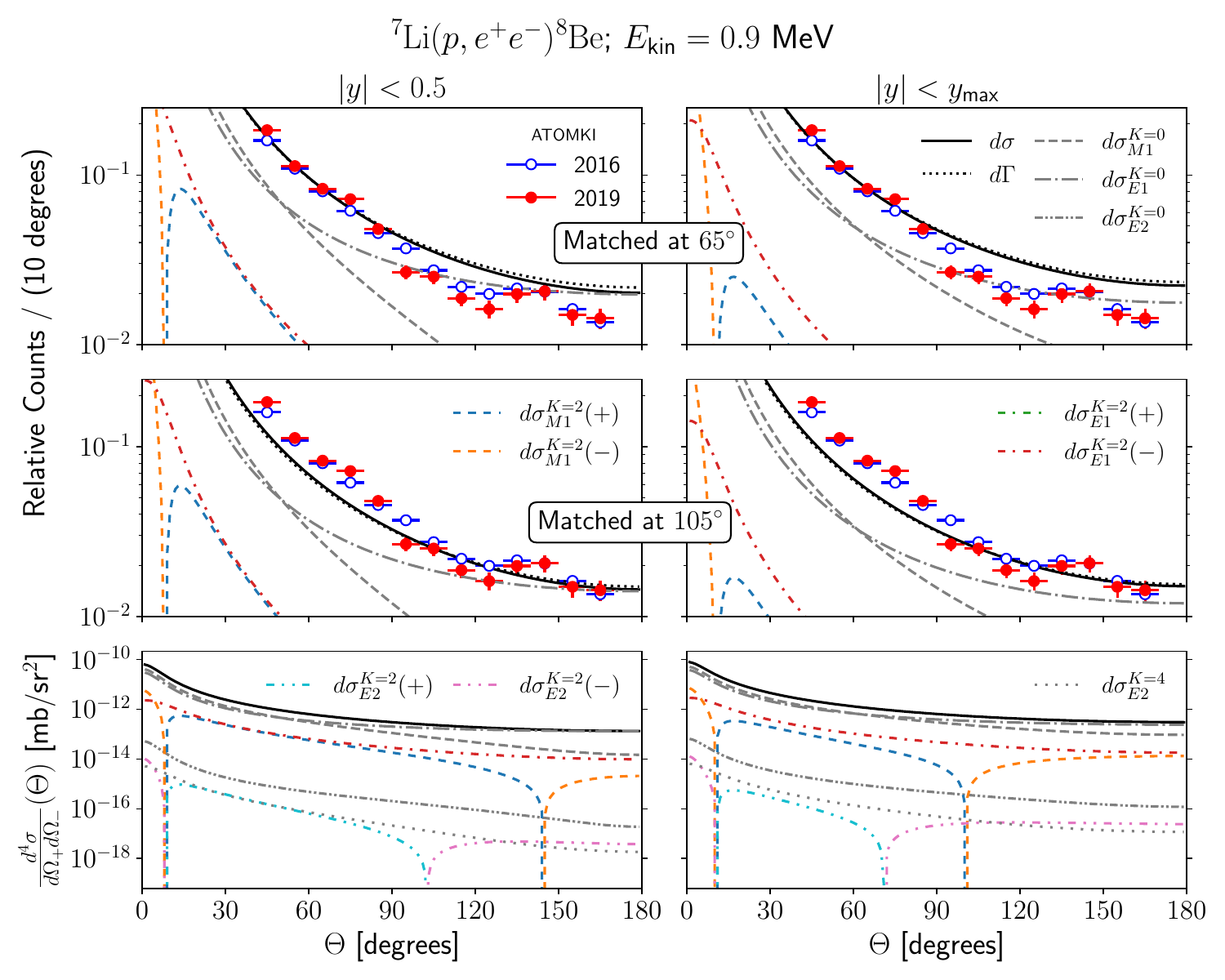}
    \caption{Calculated internal pair conversion differential cross section dependence on the angle between $e^+$ and $e^-$ (bottom panels) matched to ATOMKI data~\cite{Krasznahorkay2016,Firak2020} at $65^\circ$ (top panels) and at $105^\circ$ (middle panels). NCSMC-pheno calculations in $N_{\rm max}{=}9$ space are shown. The left panels show the cross section integrated over the energy asymmetry $y$ in the range $|y|{<}0.5$ corresponding to the experimental limitations while in the right panels the full kinematic-allowed region is used. The $d\Gamma$ refers to an approximation equivalent to $d\sigma$ including only $K{=0}$ components. See the text for further details.}
    \label{fig:pairprod_2016}
\end{figure*}
The calculated internal pair conversion differential cross section dependence on the angle between $e^+$ and $e^-$ at the $^7$Li+$p$ center-of-mass energy of the 18.15 MeV $1^+_2$ resonance is presented in Fig.~\ref{fig:pairprod_2016}. The solid lines represent the total. We present the $K{=}0$ $E1$ and $M1$ contribution by dash-dotted and dashed lines, respectively. At high angles the $E1$ component flattens out while the $M1$ keeps decreasing. In the lower panels, the $K=0$ $E2$ contribution is also visible, presented by dash-dot-dotted lines. The subleading $K{=}2$ contributions are shown by loosely dash-dotted ($E1$), loosely dashed ($M1$) and loosely dash-dot-dotted ($E2$) lines with the $(-)$ label indicating that the contribution needs to be multiplied by $-1$. Only $E2$ contributes to the $K=4$ contribution, presented by loosely dotted lines. In the top and middle panels, the differential cross section is scaled to ATOMKI data~\cite{Krasznahorkay2016,Firak2020} by matching it at the $65^\circ$ and $105^\circ$, respectively. The dotted lines in those panels represent the sum of the $K{=}0$ $E1$, $M1$ and $E2$ contributions, which is an approximation used, e.g., in Ref.~\cite{Hayes2022} and is equivalent to Eq. (\ref{eq:dGamma_pp}). Similar to Ref.~\cite{Viviani2022} we can include the interference from different partial waves in the scattering state using the full theoretical
prediction (\ref{eq:full_diff_sigma}). The left panels show the cross section integrated over the energy asymmetry $y{=}(E_+-E_-)/\omega$ (\ref{eq:y}) in the range $|y|{<}0.5$ corresponding to the experimental limitations of the measurements reported in Refs.~\cite{Krasznahorkay2016,Firak2020} while in the right panels the full kinematic-allowed region is used. We can see that in the latter case the differences between the solid and the dotted lines diminish while there are some visible differences in the former case. Clearly, including the $K{=}2$ and $K{=}4$ interference terms brings the calculation closer to the experimental data. However, overall when matched at the lower angle, the data are overestimated at high angles while when matched at high angles the data are underestimated at low angles. A similar pattern has been reported in Ref.~\cite{Hayes2022}. In that work, a phenomenological R-matrix fit to the data was used to obtain the $E1/M1$ ratio. In the present study we compute all the components within the NCSMC with our predictions supported by the calculation of radiative capture (Fig. \ref{fig:capture}). Obviously, the calculated differential cross section is smooth. The bump at ${\approx} 140^\circ$ seen in the data, if real, cannot be explained by a Standard Model electromagnetic process. Overall, our calculations are reasonably close although clearly not in a perfect agreement with the ATOMKI 2016 and 2019 electromagnetic background (i.e., without the bump) data. However, we are not in a position to make any statement about a flaw in those data. 

\begin{figure*}
    \centering
    \begin{subfigure}{0.5\textwidth}
        \includegraphics[width=\textwidth]{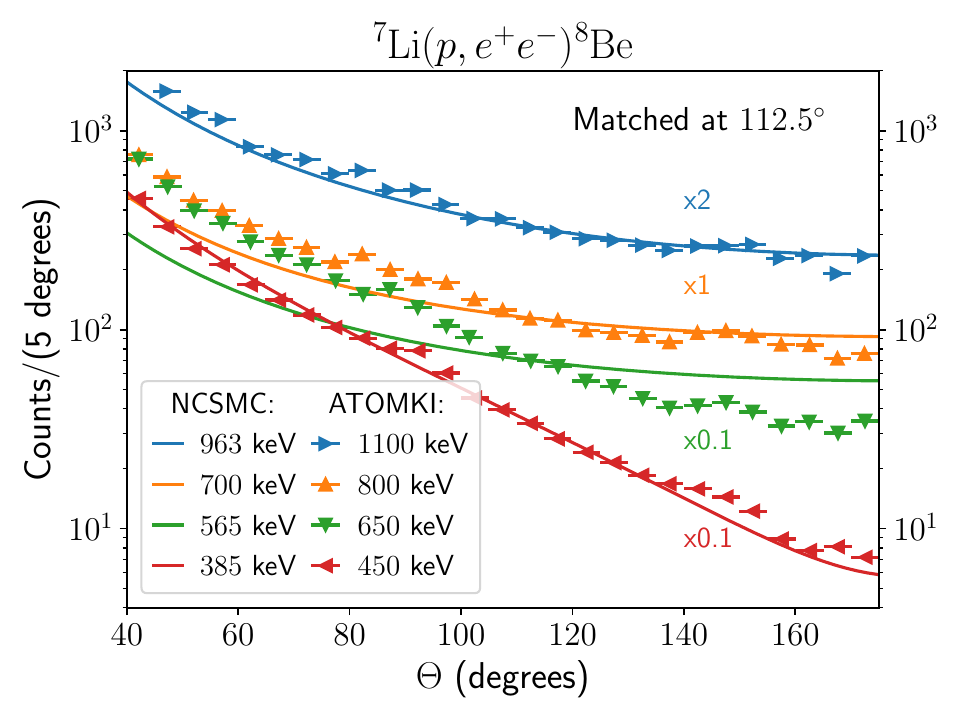}
        \caption{}
    \end{subfigure}%
    \begin{subfigure}{0.5\textwidth}
        \includegraphics[width=\textwidth]{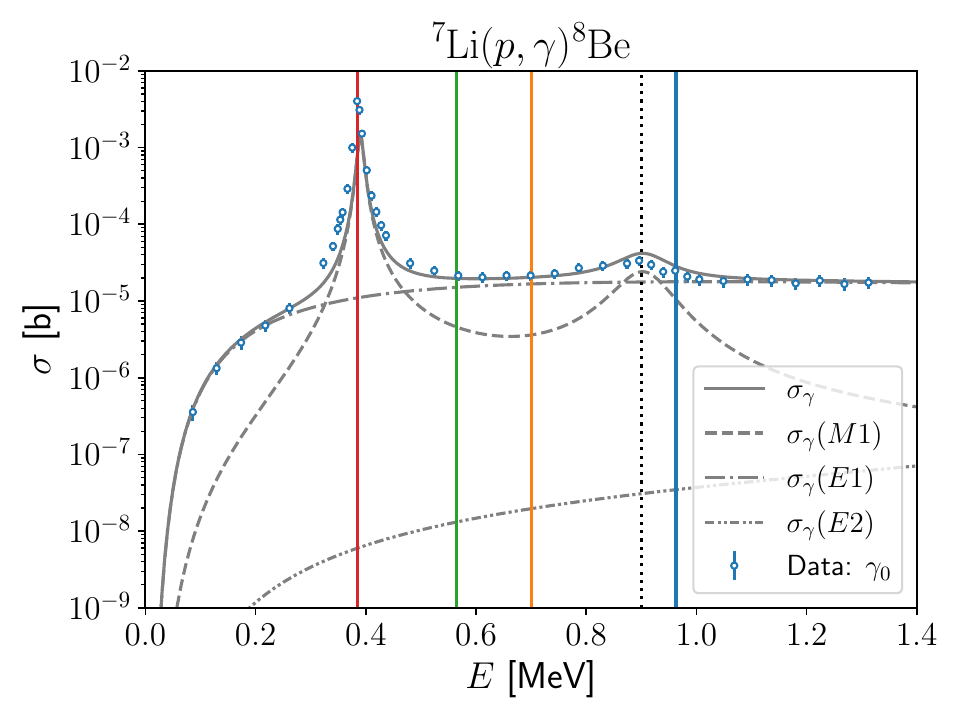}
        \caption{}
    \end{subfigure}
    \caption{(a) Calculated internal pair conversion differential cross section dependence on the angle between $e^+$ and $e^-$ matched to ATOMKI data~\cite{Sas:2022pgm} at $112.5^\circ$. The center-of-mass and the proton energies are shown at the bottom left. The cross section was integrated over the energy asymmetry $|y|{<}0.3$ corresponding to the experimental limitations reported in Ref.~\cite{Sas:2022pgm}. (b) A dependence of the integrated $\gamma$ radiative capture cross section to the $^8$Be ground state on the $^7$Li${+}p$ energy in the center-of-mass. Experimental data from Ref.~\cite{Zahnow1995}. The $M1$, $E1$, and $E2$ contributions are shown as dashed, dash-dotted and dash-dot-dotted lines respectively. The solid vertical lines indicate energies of the internal pair production measurements from the panel (a). The vertical dotted line shows the resonance centroid. The NCSMC-pheno calculations were performed in $N_{\rm max}{=}9$ space.}
    \label{fig:pairprod_2022}
\end{figure*}
Very recently the ATOMKI collaboration reported new measurements of the $^7$Li($p,e^+ e^-$)$^8$Be internal pair conversion correlations at four different proton energies; at the 17.64 MeV $1^+_1$ resonance, at ${\approx} 70$ keV above the centroid of the 18.15 MeV $1^+_2$ resonance as well as at two energies between the resonances~\cite{Sas:2022pgm}. The experimental data for several proton energies are shown in Fig.~\ref{fig:pairprod_2022} (a) and compared to our NCSMC calculations at corresponding center-of-mass energies, matched to the data at $112.5^\circ$. In the panel (b) of Fig.~\ref{fig:pairprod_2022}, the NCSMC calculated integrated $^7$Li($p,\gamma$)$^8$Be cross section is presented together with the experimental data from Ref.~\cite{Zahnow1995}. A quite reasonable agreement with the $\gamma$-capture data is found as already pointed out in Sect.~\ref{radcapture}. The solid vertical lines indicate the energies of the internal pair conversion measurements presented in the panel (a). Our NCSMC calculations reproduce quite well the $e^+ e^-$ correlation data at both resonances (i.e., 385 keV and 963 keV), in fact better than the 2016/2019 data (Fig.~\ref{fig:pairprod_2016}). However, the NCSMC results are much flatter compared to the data at energies between the resonances (565 keV and 700 keV) with the former especially off. As seen in the panel (b), the calculations in between the resonances are completely dominated by the $E1$ contribution producing a flat $e^+ e^-$ angular dependence while at or near the resonances the $M1$ contribution either dominates (at $1^+_1$ resonance) or is significant resulting in a steeper decrease with the increasing angle. It is argued in Ref.~\cite{Sas:2022pgm} that there might be contributions from protons slowed-down in the target and captured at the 17.64 MeV $1^+_1$ resonance to the 565 keV and even to the 700 keV data. If so, our calculations performed at a fixed energy cannot be directly compared to such data. This explanation appears plausible as the 565 keV data are significantly steeper, i.e., include a higher $M1$ contribution, and disagree with our calculations more than the 700 keV data further away from the $1^+_1$ resonance.

\subsection{Emission of Hypothetical Particles}

\begin{figure}[tbph]
	\centering
		\includegraphics[width=\linewidth]{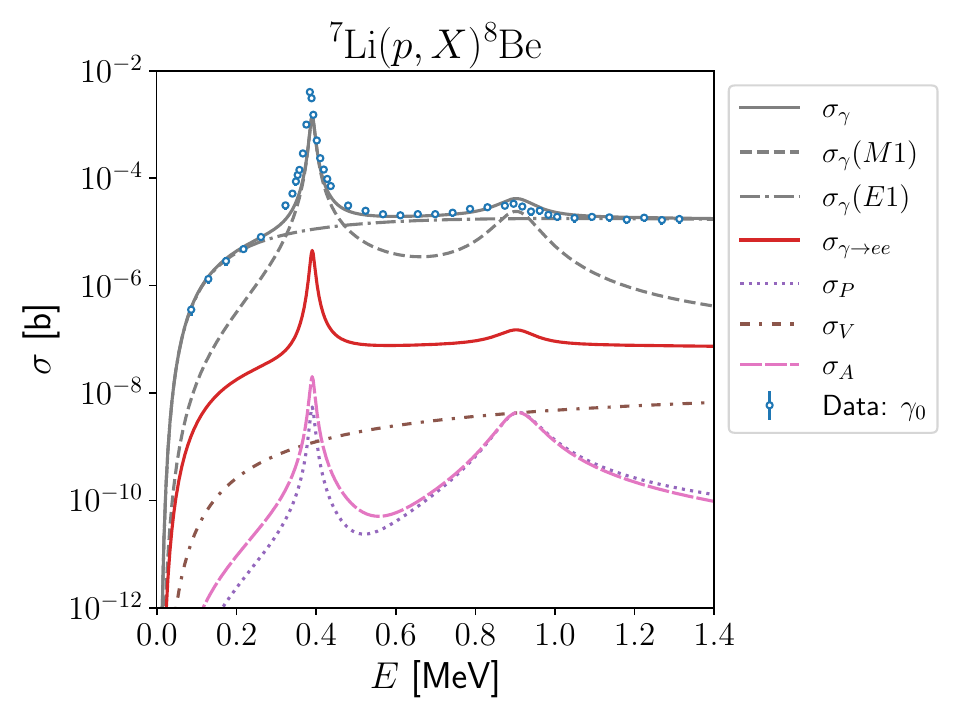}
	\caption{The integrated proton capture cross section on $^7$Li dependence on the energy in the center-of-mass. Calculated $\gamma$ emission cross section with $E1$ and $M1$ components shown separately is compared to experimental data from Ref.~\cite{Zahnow1995}, to the integrated $e^+ e^-$ internal pair conversion cross section, and to the calculated hypothetical X17 boson emissions considering the $E1$ vector, pseudoscalar, and axial vector boson candidates. See the text for further details.}
	\label{fig:capture_X}
\end{figure}

An estimate of the integrated cross section of the reaction $\ce{^{7}Li}(p,X)\ce{^{8}Be}$ was calculated with three X17 candidates: a pseudoscalar \cite{Donnelly1978} using the operator~(\ref{eq:P}), axial vector \cite{Kozaczuk2017} with the operator~(\ref{eq:A}) and vector \cite{Zhang2021} with the operator~(\ref{eq:V}). The cross section is computed by substituting the leading-order operator in the long-wavelength approximation into the formula for radiative capture (\ref{eq:radcap_cross}). The results for transitions to the $^8$Be $0^+$ ground state are shown in Figure \ref{fig:capture_X}. For the pseudoscalar and axial vector, we use the couplings from Ref.~\cite{Backens2022}, which scaled the operators based on the strength of the signal in the 2016/2019 ATOMKI data \cite{Firak2020}. For the $E1$ vector, we adjusted the coupling to the second peak of the other models. 

We also calculated the integrated $\ce{^{7}Li}(p,e^{+}e^{-})\ce{^{8}Be}$ cross section. Technical details of the integration are given in Appendix~\ref{sec:pairprod_radcap_appendix}.
As seen in Fig.~\ref{fig:capture_X}, it is essentially proportional to the $\gamma$-emitting proton-capture cross section with its magnitude scaled by a factor of ${\approx} \alpha/\pi{\approx} 10^{-3}$. We note that the interference between $\gamma$ and $X$ is not included in the present calculations. It is understandable
that an anomaly would be hardly observed at the first resonance which is
swamped by the very high electromagnetic $M1$ rate. The effect from
the hypothetical X17 boson could be expected at second resonance in
the case of a pseudoscalar, axial vector, or vector character. Any anomaly
present between the resonances and at the second resonance could be consistent with a vector particle, which is the preferred candidate in the most recent ATOMKI publications
\cite{Sas:2022pgm,Krasznahorkay2022}. 

\section{Conclusions}\label{concl}

We applied {\it ab initio} NCSMC with chiral NN+3N interactions as input to investigate the structure of $^8$Be, proton-$^7$Li elastic scattering, $^7$Li($p,\gamma$)$^8$Be radiative capture and the internal pair creation $^7$Li($p,e^+ e^-$)$^8$Be. We included the $^7$Li${+}p$ and $^7$Be${+}n$ clusters in the basis which allowed us to treat properly the isospin breaking effects and study the physics of $1^+$ resonances that appear in-between the thresholds of the two mass partitions. Contrary to other methods that have been applied to calculate the $^8$Be decays so far, within the NCSMC we are able to compute in a unified framework both resonant and non-resonant $e^+e^-$ pair production with realistic isospin-breaking effects, mediated by a virtual photon or by X17.

Overall, we demonstrated a good description of $^8$Be states both below the $^7$Li${+}p$ threshold as well as low-lying (up to a few MeV) resonances above this threshold. We were able to reproduce existing data for the $^7$Li($p,\gamma$)$^8$Be radiative capture integrated cross section. To facilitate the comparison to experimental data, we applied phenomenological corrections to reproduce experimental thresholds and positions of the lowest two $1^+$ resonances.

After validating our approach as described above, we calculated internal pair conversion differential cross section dependence on the angle between $e^+$ and $e^-$ and compared them to available data sets by ATOMKI collaboration. As expected, we obtained smooth differential cross sections, i.e., without any bump, dominated by $M1$ contribution at and close to resonances and by $E1$ contribution away from resonances. Comparing to ATOMKI 2016/2019 data sets taken at the 18.15 MeV $1^+$ resonance~\cite{Krasznahorkay2016,Firak2020}, we observe that by matching our calculation to the experimental data at a lower angle, the data are somewhat overestimated at high angles while when matched at a high angle the data are somewhat under-estimated at low angles. We have shown that including interference terms that appear due to the treatment of the initial state as a scattering state brings the calculation closer to the experimental data. 

We also compared our internal pair conversion differential cross section calculations to the ATOMKI 2022 data sets measured at four energies, at the 17.64 MeV $1^+$ resonance, at ${\approx} 70$ keV above the centroid of the 18.15 MeV $1^+$ resonance as well as at two energies between the resonances~\cite{Sas:2022pgm}. Our NCSMC calculations reproduce quite well the $e^+ e^-$ angular correlation data at both resonances, in fact better than the 2016/2019 data. However, the NCSMC results are much flatter compared to the data at energies between the resonances due to a complete dominance by the $E1$ contribution producing a flat $e^+ e^-$ angular dependence. It is argued in Ref.~\cite{Sas:2022pgm} that there might be contributions from protons slowed-down in the target and captured at the 17.64 MeV $1^+$ resonance to the between-resonances data. If so, our calculations performed at a fixed energy cannot be directly compared to such data.

Finally, we performed integrated cross section calculations for three model candidates for the hypothetical X17 particle; a pseudoscalar, axial vector, and vector. We showed that the effect of the hypothetical boson could be expected at 18.15 MeV $1^+$ resonance in the case of a pseudoscalar, axial vector, or vector character while a signal at the 17.64 MeV $1^+$ resonance would likely be overwhelmed by the isovector $M1$ Standard Model $e^+ e^-$ background. An anomaly present between the resonances and at the second resonance could be consistent with a vector particle. The latter is the preferred candidate according to the latest ATOMKI publications~\cite{Sas:2022pgm,Krasznahorkay2022}.

Theoretical calculations like the ones we embarked on and presented in this paper can point at inconsistencies in experimental data. However, they cannot answer the question if the X17 anomaly is real and the proposed new boson exists. Our results are to a large extent in line with the ATOMKI Standard Model background, i.e., the $e^+ e^-$ angular correlations without the anomaly bump at ${\approx} 140^\circ$, although we call for a careful analysis of the apparent contamination of the data between the resonances due to the proton energy loss in the thick target. Clearly, independent experimental measurements are needed to confirm or dispute the existence of the X17 anomaly. There are several experiments under way or in preparation. For example, the NewJEDI~\cite{NewJedi2023} collaboration already performed independent $^7$Li($p,e^+ e^-$)$^8$Be measurements and is currently analysing their data.

Our calculations can be improved by including the $^4$He${+}^4$He mass partition. We expect, however, that it would have a very limited impact on the capture reactions to the $^8$Be ground state studied here. We plan to study the interference of photon and the hypothetical X17 boson in the the $e^+ e^-$ angular correlations following the approach of Ref.~\cite{Viviani2022}. Also, we will further investigate the higher lying resonances in  $^8$Be that our calculations predict and study charge exchange reactions $^7$Li($p,n$)$^7$Be and $^7$Be($n,p$)$^7$Li. Finally, we will calculate the $^{11}$B($p,e^+ e^-$)$^{12}$C angular correlations recently measured by the ATOMKI collaboration~\cite{Krasznahorkay2022}.  

\section{Acknowledgements}
We thank Michele Viviani for in-depth discussions of the internal $e^+ e^-$ pair production formalism. We also thank Chloe Hebborn for discussion of radiative capture calculations and Kang Choi for discussions of pair production. This work was supported by the NSERC Grants No. SAPIN-2022-00019, SAPIN-2016-00033 and PGSD3-535536-2019 and by the U.S. Department of Energy, Office of Science, Office of Nuclear Physics, under Work Proposals No. SCW0498. TRIUMF receives federal funding via a contribution agreement with the National Research Council of Canada. This work was prepared in part by LLNL under Contract No. DE-AC52-07NA27344. Computing support came from an INCITE Award on the Summit supercomputer of the Oak Ridge Leadership Computing Facility (OLCF) at ORNL, from the LLNL institutional Computing Grand Challenge program, and the Digital Research Alliance of Canada.

\appendix

\begin{widetext}
\section{Nuclear Transition Matrix Elements of Electromagnetic Multipole Operators}
\label{sec:multipole_operators}

Here and in the following appendices, we use natural units $\hbar=c=1$ such that the energy and momentum use the same units, and the momentum is equivalent to the wavenumber.

The interaction of a nucleus with an oscillating electromagnetic field corresponds to the Fourier transform of the charge $\rho$ and current $\vec{\mathcal{J}}$ density operators, together the four-vector $\mathcal{J}_\mu=(\rho,\vec{\mathcal{J}})$, i.e.
\begin{align}
	\rho(\vec{q}) &= \int \dd^3 r e^{-i\vec{q}\cdot\vec{r}} \rho(\vec{r}) \;, \\
	\vec{\mathcal{J}}(\vec{q}) &= \int \dd^3 r e^{-i\vec{q}\cdot\vec{r}} \vec{\mathcal{J}}(\vec{r}) \;,
\end{align}
where, with $m_p$ the proton mass, \cite{Akhiezer, Walecka}
\begin{align}
	\rho(\vec{r}) &= \sum_{i=1}^A \frac{1+\tau_{zi}}{2} \delta(\vec{r}-\vec{r}_i) \;, \\
	\vec{\mathcal{J}}(\vec{r}) &= \sum_{i=1}^A \frac{1+\tau_{zi}}{2} \frac{\vec{\nabla}\delta(\vec{r}-\vec{r}_i) + \delta(\vec{r}-\vec{r}_i)\vec{\nabla}}{2im_p} + \frac{1}{2m_p}\sum_{i=1}^A g_{si} \frac{\vec{\nabla}\times\vec{S}_i\delta(\vec{r}-\vec{r_i}) + \delta(\vec{r}-\vec{r_i})\vec{\nabla}\times\vec{S}_i}{2} \;.
\end{align}
A standard technique \cite{Donnelly1975, deForest1966, Walecka} is to expand the vector current in terms of the spherical unit vectors $\vec{e}_\lambda$, $\lambda\in \{-1,0,1\}$, i.e.
\begin{equation}
	\vec{\mathcal{J}}(\vec{q}) = \sum_\lambda \mathcal{J}_\lambda(\vec{q}) \vec{e}^*_\lambda \;.\label{eq:J_expansion}
\end{equation}
The components correspond to the photon polarizations: longitudinal $\vec{e}_0=\hat{z}$ and transverse $\vec{e}_{\pm1} = \mp \frac{1}{\sqrt{2}}(\vec{x} \pm i\vec{y})$, when we choose a set of coordinate axes ($\hat{x}$, $\hat{y}$, $\hat{z}$) such that $\vec{q} = q \hat{z}$.
The spherical unit vectors have the properties $\vec{e}_\lambda \cdot\vec{e}_{\lambda'}^* = \delta_{\lambda\lambda'}$ and $\vec{e}_\lambda^* = (-1)^\lambda \vec{e}_{-\lambda}$.

Making use of the spherical Bessel functions $j_J(qr)$, the spherical harmonics $Y_{JM}(\theta,\phi)$ and vector spherical harmonics $\vec{Y}_{JLM}(\theta,\phi)=\sum_{m\lambda} (JM1\lambda|Lm) Y_{LM}(\theta,\phi) \vec{e}_\lambda $, we may expand in terms of multipole operators.
With the combinations $M_{JM}(q) = j_J(qr) Y_{JM}(\Omega_r)$ and $\vec{M}_{JLM}(q) = j_J(qr) \vec{Y}_{JLM}(\Omega_r)$, we may define four types of multipole operators:
\begin{itemize}
	\item Coulomb:
	\begin{equation}
		\mathcal{C}_{JM}(q) = \int \dd^3 r M_{JM}(q,\vec{r})\rho(\vec{r}) \label{eq:C}
	\end{equation}
	\item Longitudinal:
	\begin{equation}
		\mathcal{L}_{JM}(q) = \int \dd^3 r \left( \frac{i\vec{\nabla}}{q} M_{JM}(q,\vec{r})\right)\cdot\vec{\mathcal{J}}(\vec{r}) \label{eq:L}
	\end{equation}
	\item Transverse Electric:
	\begin{equation}
		\mathcal{T}^E_{JM}(q) = \int \dd^3 r \left(\frac{\vec{\nabla}}{q}\times \vec{M}_{JJM}(q,\vec{r})\right)\cdot \vec{\mathcal{J}}(\vec{r}) \label{eq:T_E}
	\end{equation}
	\item Transverse Magnetic:
	\begin{equation}
		\mathcal{T}^M_{JM}(q) = \int \dd^3 r \vec{M}_{JJM}(q,\vec{r})\cdot \vec{\mathcal{J}}(\vec{r}) \label{eq:T_M}
	\end{equation}
\end{itemize}
Hence, the charge and current operators may be expressed as
\begin{align}
	\rho(q) &= \sqrt{4\pi} \sum_{j\geq 0} (-i)^j \hat{j} \mathcal{C}_{j0}(q) \;, \\
	\mathcal{J}_\lambda(q) &= 
	\begin{cases}
		\sqrt{4\pi} \sum_{j\geq 0} (-i)^j \hat{j} \mathcal{L}_{j0}(q) & \text{if } \lambda=0 \;,\\
		\sqrt{2\pi} \sum_{j\geq 1} (-i)^j \hat{j} \left(\mathcal{T}^E_{j\lambda}(q) -\lambda\mathcal{T}^M_{j\lambda}(q)\right) & \text{if } \lambda=\pm 1 \;.
	\end{cases}
\end{align}

Following \cite{Walecka}, partial integration may be used to move gradients to apply to $\vec{\mathcal{J}}$ and various vector identities may be used to rearrange the expressions.
Through the conservation of the charged current $\vec{\nabla}\cdot\vec{\mathcal{J}}(q)=-\frac{\dd \rho}{\dd t}=-i\left[ H, \rho(q) \right]$, evaluated between eigenstates of the Hamiltonian, $\bra{f}$ and $\ket{i}$, longitudinal multipoles may be replaced in favour of Coulomb multipoles, i.e.
\begin{equation}
	\mathcal{L}_{JM}(q) = -\frac{E_f-E_i}{q} \mathcal{C}_{JM}(q) = \frac{\omega}{q} \mathcal{C}_{JM}(q)\;.
	\label{eq:Siegert}
\end{equation}
In photon emission $\omega=E_i-E_f$ takes a positive value. Equivalently $\vec{q}\cdot \vec{\mathcal{J}}(q)=0$, hence $\mathcal{J}_0(q) = \frac{\omega}{q}\rho(q)$.

Further, at low-energy ($qr\ll1$), we may Taylor expand the Bessel functions, i.e. 
\begin{equation}
	j_J(qr) = \frac{(qr)^J}{(2J+1)!!}\left( 1 - \frac{\tfrac{1}{2}(qr)^2}{2J+3} \cdots \right) \;.
\end{equation}
This long-wavelength approximation allows us to use the standard spatial multipole operators (\ref{eq:M_E}, \ref{eq:M_M}), i.e.
\begin{align}
	\mathcal{C}_{JM}(q) &\simeq \frac{q^J}{(2J+1)!!} \mathcal{M}^E_{JM} \;, \\
	\mathcal{L}_{JM}(q) &\simeq \frac{\omega}{q} \mathcal{C}_{JM}(q) \simeq \frac{\omega q^{j-1}}{(2j+1)!!}\mathcal{M}^E_{JM} \;, \\
	\mathcal{T}^E_{JM}(q) &\simeq \sqrt{\frac{J+1}{J}}\mathcal{L}_{JM}(q) \simeq \frac{\omega q^{J-1}}{(2J+1)!!}\sqrt{\frac{J+1}{J}}\mathcal{M}^E_{JM} \;, \\
	\mathcal{T}^M_{JM}(q) &\simeq \frac{iq^J}{(2J+1)!!} \sqrt{\frac{J+1}{J}} \mathcal{M}^M_{JM} \label{eq:longwave_approx}
\end{align}
where
\begin{align}
	\mathcal{M}^E_{jm} &= \sum_{i=1}^A \frac{1+\tau_{zi}}{2} r_i^j Y_{jm} (\Omega_i) \;, \\
	\mathcal{M}^M_{jm} &= \frac{1}{2m_p} \sum_{i=1}^A \left( \frac{2g_{\ell i}}{j+1} \vec{L}_i + g_{si} \vec{S}_i \right) \cdot \vec{\nabla}_i \left( r_i^j  Y_{jm} (\Omega_i) \right) \;.
\end{align}
The difference between (\ref{eq:M_E}, \ref{eq:M_M}) in the main text and the above is the multiplication by the elementary charge $e$ (the nuclear magneton $\mu_N$ is defined by $\frac{e}{2m_p}$). In these appendices we explicitly multiply $\mathcal{J}_\mu$ by $e$, as $e$ also appears in the photon-lepton interactions.
The elementary charge may be defined as $e=\sqrt{4\pi\alpha\epsilon_0\hbar c}$, in natural units $e^2 = 4\pi\alpha$.

\section{Extended Derivation - Radiative Capture}
\label{sec:radcap_appendix}

In a radiative capture reaction $T(P,\gamma)F$, as shown in \autoref{fig:feyn_cap}, we calculate the
cross section for a $2\to2$ process \cite{Griffiths_PP}, i.e.
\begin{align}
	\dd\sigma= & \frac{1}{v}\frac{1}{4E_{P}E_{T}}\bar{\sum_{i}}\sum_{f}\sum_{\lambda}\left|\mathcal{M}_{FI}^{\lambda Q}\right|^{2}\frac{\dd^{3}q}{(2\pi)^{3}2\omega}\frac{\dd^{3}p_{F}}{(2\pi)^{3}2E_{F}}\;.
\end{align}
%
The initial state consists of two nuclei: the target $T$ and projectile $P$. The average
over initial states is an average over the angular momenta of both
clusters (target $\ket{s_{T}m_{T}}$ and projectile $\ket{s_{P}m_{P}}$), i.e. $\bar{\sum_{i}}=\frac{1}{\hat{s}_{P}^{2}\hat{s}_{T}^{2}}\sum_{m_{P}m_{T}}$.
The sum over final states $\ket{J_{f}M_{f}}$ is $\sum_{f}=\sum_{M_{f}}$.

The squared amplitude is
\begin{align}
	\left|\mathcal{M}_{FI}^{\lambda Q}\right|^2= & (2\pi)^{4}\delta(P_{P}+P_{T}-P_{F}-Q)e^2\left|H_{\lambda}^{fi}\right|^{2}8E_{F}E_{P}E_{T} \;.
\end{align}
where
\begin{align}
	H_{\lambda}^{fi}= & \bra{f}(-\vec{e}_{\lambda}^{*}\cdot\vec{\mathcal{J}}(\vec{q}))\ket{i} \label{eq:H_fi_lambda}\;.
\end{align}
Cancelling the $2E$ factors and integrating over the final state momentum, the
cross section is then
\begin{align}
	\dd\sigma= & \frac{e^{2}}{v}\frac{\dd^{3}q}{(2\pi)^{3}2\omega}2\pi\delta(E_{P}+E_{T}-E_{F}-\omega)\bar{\sum_{i}}\sum_{f}\sum_{\lambda}\left|H_{\lambda}^{fi}\right|^{2}\label{eq:radcap_ds_Hl}\;.
\end{align}

Implicit in the definition of the multipole operators is that the
$\lambda=\pm1$ polarizations are transverse to $\vec{q}$, the direction
of the photon. However the projection of the spherical tensor operator
should be defined in the same coordinates as the quantum states it
acts on. In a reaction, we define the angular momentum projection of
the nuclear states with respect to the relative velocity (i.e. the direction
of the beam). This requires us to rotate the operators (or states).
The rotation operator can be written in terms of the total angular momentum operators $J_z$ and $J_y$, following the convention of Ref. \cite{Varshalovich} i.e. 
\begin{equation}
	\mathcal{R}(\alpha,\beta,\gamma)=e^{-i\alpha J_z}e^{-i\beta J_y}e^{-i\gamma J_z}\;.
\end{equation}
The matrix elements of the rotation operator between eigenstates of the angular momentum are the Wigner D-matrices, i.e.
\begin{equation}
	\bra{j'm'}\mathcal{R}(\alpha,\beta,\gamma)\ket{jm} = \delta_{j'j} D^j_{m'm}(\alpha,\beta,\gamma)
\end{equation}
where
\begin{equation}
	D^j_{m'm}(\alpha,\beta,\gamma) = e^{-im'\alpha} d^j_{m'm}(\beta) e^{-im\gamma} \;.\label{eq:D_mat}
\end{equation}
%

The set of coordinates defined relative to $\hat{z}$ along the beam is the ``$\hat{v}$-frame'', while the set of coordinates defined relative to $\hat{z}$ along the photon is the ``$\hat{q}$-frame''.
In our case, the direction of the photon is described by the polar coordinates $(\theta_q,\phi_q)$ in the $\hat{v}$-frame.
The corresponding rotation $\hat{v}\to \hat{q}$ is $\mathcal{R}(\phi_q,\theta_q,-\phi_q)$ \cite{JacobWick}.
The state $\ket{i}$ in the $\hat{q}$-frame is $\ket{i}_q = \mathcal{R}(\phi_q,\theta_q,-\phi_q)\ket{i}_v$, related by a rotation operator to the state $\ket{i}$ in the $\hat{v}$-frame.
The matrix element is expanded over multipole operators via (\ref{eq:eJ_1}) in the $\hat{q}$-frame. In order to use states in the $\hat{v}$-frame we apply the rotation operators. Using properties of the rotation operators, i.e., the conjugate of $\mathcal{R}$, equivalent to the inverse, implies rotation around each axis in the opposite direction and in the opposite order,
\begin{equation*}
  \mathcal{R}^\dagger(\alpha,\beta,\gamma)=\mathcal{R}(-\gamma,-\beta,-\alpha)\;,
\end{equation*}
and a property of spherical tensor operators
\begin{equation*}
  \mathcal{R}\mathcal{O}_{jm}\mathcal{R}^\dagger=\sum_{m'}\mathcal{O}_{jm'}D^{j}_{m'm}\;,
\end{equation*}
the result contains the $D$-matrices, i.e.
\begin{align}
	\bra{f}_q (-\vec{e}_\lambda^* \cdot \vec{\mathcal{J}}(q))\ket{i}_q =& \bra{f}_v \mathcal{R}^\dagger (\phi_q,\theta_q,-\phi_q) (-\vec{e}_\lambda^* \cdot \vec{\mathcal{J}}(q)) \mathcal{R}(\phi_q,\theta_q,-\phi_q) \ket{i}_v \nonumber \\
	=& \bra{f}_v \sum_{jm}(-i)^{j}\sqrt{2\pi}\hat{j}\left[\mathcal{T}_{jm}^{E}(q)+\lambda\mathcal{T}_{jm}^{M}(q)\right]D_{m-\lambda}^{j}(-\phi_q,-\theta_q,\phi_q) \ket{i}_v\;.
\end{align}
In the following manipulations we use $\ket{i}=\ket{i}_v$ and so we insert into (\ref{eq:H_fi_lambda}) the operator:
\begin{align}
	-\vec{e}_{\lambda}^{*}\cdot\vec{\mathcal{J}}(q)= & \sum_{jm}(-i)^{j}\sqrt{2\pi}\hat{j}\left[\mathcal{T}_{jm}^{E}(q)+\lambda\mathcal{T}_{jm}^{M}(q)\right]D_{m-\lambda}^{j}(-\phi_q,-\theta_q,\phi_q)\;.
\end{align}

The initial continuum state in radiative capture can be expanded over a coupled basis, the states of (\ref{eq:cluster_state}), i.e.
\begin{align}
	\ket{\Psi_{\nu_i}^{(m_Tm_P)}}= & \frac{\sqrt{4\pi}}{k_{\nu_i}}\sum_{J^\pi SL}i^L\hat{L}e^{i\sigma_{L}}\left(s_{T}m_{T}s_{P}m_{P}|SM\right)\left(SML0|JM\right)\ket{\Psi_{\nu_{i}SL}^{J^\pi M}}\;,\label{eq:initial_states}
\end{align}
where $\nu_i$ describes the mass partition and internal states ($\nu_T$, $\nu_P$). The symbol $M$ is fixed to the value $m_T+m_P$.
The state $\Psi_{\nu_i SL}^{J^\pi M}$ corresponds to the NCSMC solution which at large cluster separations (where the composite NCSM part of the wave function and the inter-cluster antisymmetrization effects are negligible) can be approximated by 
\begin{align}
	\ket{\Psi_{\nu_i SL}^{J^\pi M}} = \sqrt{v_{\nu_i}}\sum_{\nu s\ell} \frac{\chi^{J^\pi}_{\nu s \ell, \nu_i SL}(r_\nu)}{r_\nu} \left[ \left[ \ket{(A-a)\nu_T s_T^{\pi_T}}\ket{a \nu_Ps_P^{\pi_P}} \right] ^{(s)} Y_\ell(\hat{r}_\nu) \right] ^{(J^\pi )}_{M} \label{eq:initial_state_internal}
\end{align}
with $\chi^{J\pi}$ defined consistently with Eq. (35) of \cite{Baroni2013}.

The sum in (\ref{eq:radcap_ds_Hl}) can be expanded
\begin{align}
	\bar{\sum_{i}}\sum_{f}\sum_{\lambda}\left|H_{\lambda}^{fi}\right|^{2}= & \frac{1}{\hat{s}_{T}^{2}\hat{s}_{P}^{2}}\sum_{m_{T}m_{P}}\sum_{M_{f}}\sum_{\lambda}\left(\bra{J_{f}M_{f}}\sum_{\kappa jm}D_{m-\lambda}^{j}\left[ \sqrt{2\pi}\hat{j}\lambda^{\kappa}(-i)^{j}\mathcal{T}_{jm}^{\kappa}\right]\ket{s_{T}m_{T}s_{P}m_{P}}\right)\nonumber \\
	& \times\left(\bra{J_{f}M_{f}}\sum_{\kappa'j'm'}D_{m'-\lambda}^{j'}\left[\sqrt{2\pi}\hat{j}'\lambda^{\kappa'}(-i)^{j'}\mathcal{T}_{j'm'}^{\kappa'}\right]\ket{s_{T}m_{T}s_{P}m_{P}}\right)^{*}\;,
\end{align}
where $\kappa=(0,1)$ corresponds to (E,M) respectively. The angles $\theta_q$, $\phi_{q}$ are suppressed as well as the $\nu_i$ labels, i.e. $\ket{s_{T}m_{T}s_{P}m_{P}}\equiv\ket{\Psi_{\nu_i}^{( m_{T}m_{P})}}$.

First, we substitute (\ref{eq:initial_states}) ($\ket{SLJM}\equiv\ket{\Psi_{\nu_i SL}^{J^\pi M}}$), and make the replacement: $X_{jm}^{\kappa}=\frac{\sqrt{4\pi}}{k}\sqrt{2\pi}\hat{j}(-i)^{j}\mathcal{T}_{jm}^{\kappa}$, i.e.
\begin{align}
	\hat{s}_{T}^{2}\hat{s}_{P}^{2}\bar{\sum_{i}}\sum_{f}\sum_{\lambda}\left|H_{\lambda}^{fi}\right|^{2}= & \sum_{m_{T}m_{P}}\sum_{M_{f}}\sum_{\lambda}\bigg\{\bra{J_{f}M_{f}}\sum_{\kappa jm}D_{m-\lambda}^{j}\lambda^{\kappa}X_{jm}^{\kappa}\nonumber \\
	& \times \sum_{LSJ}\hat{L}e^{i\sigma_{L}}\left(s_{T}m_{T}s_{P}m_{P}|SM\right)\left(SML0|JM\right)\ket{SLJM}\bigg\}\nonumber \\
	& \times\bigg\{\bra{J_{f}M_{f}}\sum_{\kappa'j'm'}D_{m'-\lambda}^{j'}\lambda^{\kappa'}X_{j'm'}^{\kappa'}\nonumber \\
	& \times \sum_{L'S'J'}\hat{L}'e^{i\sigma_{L'}}\left(s_{T}m_{T}s_{P}m_{P}|S'M'\right)\left(S'M'L'0|J'M'\right)\ket{S'L'J'M'}\bigg\}^{*}\;.
\end{align}
We make use of the sum over $m_{T},m_{P}$, i.e.
\begin{align}
	\sum_{m_{T}m_{P}}\left(s_{T}m_{T}s_{P}m_{P}|SM\right)\left(s_{T}m_{T}s_{P}m_{P}|S'M'\right)= & \delta_{SS'}\delta_{MM'}\;,
\end{align}
and the Wigner-Eckart theorem:
\begin{align}
	\bra{J_{f}M_{f}}X_{jm}^{\kappa}\ket{SLJM}= & \hat{J}_{f}^{-1}\left(JMjm|J_{f}M_{f}\right)\bra{J_{f}}\left|X_{j}^{\kappa}\right|\ket{SLJ}\;.
\end{align}
Defining $X_{J_{f}SLJ}^{\kappa j} {\equiv} \bra{J_{f}}\left|X_{j}^{\kappa}\right|\ket{SLJ} $ and making use of the fact that a product of $D$ and $D^*$ is a $D$ matrix,
\begin{align*}
  D_{mk}^{j}D_{m'k'}^{j'*}=&
  \sum_{J=\left|j-j'\right|}^{j+j'}\left(-\right)^{m'-k'}\left(jmj'-m'|J(m-m')\right)\\&\times\left(jkj'-k'|J(k-k')\right)D_{(m-m')(k-k')}^{J}  \; ,
  \end{align*}
and in the present case $m=m'$, we have
\begin{align}
	\hat{s}_{T}^{2}\hat{s}_{P}^{2}\bar{\sum_{i}}\sum_{f}\sum_{\lambda}\left|H_{\lambda}^{fi}\right|^{2}= & \hat{J}_{f}^{-2}\sum_{M_{f}}\sum_{\lambda}\sum_{\kappa jm}\sum_{LSJM}\sum_{\kappa'j'}\sum_{L'J'}i^{L-L'}\hat{L}\hat{L}'e^{i\left(\sigma_{L}-\sigma_{L'}\right)}\lambda^{\kappa+\kappa'}\nonumber \\
	& \times X_{J_{f}SLJ}^{\kappa j}X_{J_{f}SL'J'}^{\kappa'j'*}\sum_{K}(-)^{m+\lambda}\left(jmj'-m|K0\right)\left(j-\lambda j'\lambda|K0\right)D_{00}^{K}\nonumber \\
	& \times  \left(SML0|JM\right)\left(SML'0|J'M\right)\left(JMjm|J_{f}M_{f}\right)\left(J'Mj'm|J_{f}M_{f}\right)\;.
\end{align}

We identify $D_{00}^{K}(-\phi_q,-\theta_q,\phi_q)$ as the Legendre polynomial $P_{K}(\cos\theta_q)$.
The sums over $m$ and $M$ produce $6j$s by making use repeatedly of 
\begin{equation*}
  \left(j_{1}m_{1}j_{2}m_{2}|jm\right)=\left(-\right)^{j_{1}-m_{1}}\frac{\hat{j}}{\hat{j}_{2}}\left(jmj_{1}-m_{1}|j_{2}m_{2}\right) \; ,
\end{equation*}
and 
 \begin{align*}
  \left(-\right)^{j_{1}+j_{2}+j_{3}+j}\hat{j}_{12}\hat{j}_{23}\left\{
  \begin{array}{ccc} j_{1} & j_{2} & j_{12}\\ j_{3} & j & j_{23}
  \end{array}\right\} \left(j_{1}m_{1}j_{23}m_{23}|jm\right)=&\nonumber\\
  \sum_{m_{2}}\left(j_{1}m_{1}j_{2}m_{2}|j_{12}m_{12}\right)\left(j_{12}m_{12}j_{3}m_{3}|jm\right)\left(j_{2}m_{2}j_{3}m_{3}|j_{23}m_{23}\right) \; ,
  \end{align*} 
i.e.
\begin{align}
	& \sum_{M_fMm}(-)^{m}\left(jmj'-m|K0\right)\left(SML0|JM\right)\left(SML'0|J'M\right)\left(JMjm|J_{f}M_{f}\right)\left(J'Mj'm|J_{f}M_{f}\right)\nonumber \\
	= & \left(-\right)^{S-J_{f}-j-j'+K-J+J'}\hat{J}_{f}^{2}\hat{J}\hat{J}'\left\{ \begin{array}{ccc}
		J & J' & K\\
		j' & j & J_{f}
	\end{array}\right\} \left\{ \begin{array}{ccc}
		J & J' & K\\
		L' & L & S
	\end{array}\right\} \left(L0L'0|K0\right)\;,
\end{align}
where the sum over angular momentum projections is limited by $M+m=M_f$.
Then converting $X$ back to $\mathcal{T}$ (using the shorthand $\mathcal{T}_{J_{f}SLJ}^{\kappa j}{\equiv}\bra{J_{f}}\left|\mathcal{T}_{j}^{\kappa}\right|\ket{SL J}$), we have
\begin{align}
	\hat{s}_{T}^{2}\hat{s}_{P}^{2}\bar{\sum_{i}}\sum_{f}\sum_{\lambda}\left|H_{\lambda}^{fi}\right|^{2}= & \sum_{\lambda}\sum_{\kappa j}\sum_{LSJ}\sum_{\kappa'j'}\sum_{L'J'}i^{L-L'}\hat{L}\hat{L}'e^{i\left(\sigma_{L}-\sigma_{L'}\right)}\lambda^{\kappa+\kappa'}X_{J_{f}SLJ}^{\kappa j}X_{J_{f}SL'J'}^{\kappa'j'*}\nonumber \\
	& \times\sum_{K}P_{K}\left(-\right)^{S-J_{f}-j-j'+K-J+J'+\lambda}\hat{J}\hat{J}'\left\{ \begin{array}{ccc}
		J & J' & K\\
		j' & j & J_{f}
	\end{array}\right\} \left\{ \begin{array}{ccc}
		J & J' & K\\
		L' & L & S
	\end{array}\right\} \nonumber \\
	&\times \left(L0L'0|K0\right)\left(j-\lambda j'\lambda|K0\right)\nonumber \\
	= & \frac{8\pi^{2}}{k^{2}}\sum_{\kappa j\kappa'j'}\sum_{SLL'JJ'}\left(\sum_{\lambda}(-)^{\lambda}\left(j-\lambda j'\lambda|K0\right)\lambda^{\kappa+\kappa'}\right)\nonumber \\
	& \times i^{L-L'+j-j'}\hat{L}\hat{L}'e^{i\left(\sigma_{L}-\sigma_{L'}\right)}\mathcal{T}_{J_{f}SLJ}^{\kappa j}\mathcal{T}_{J_{f}SL'J'}^{\kappa'j'*}\sum_{K}P_{K}\nonumber \\
	& \times\left(-\right)^{S-J_{f}-L+L'-J+J'}\hat{j}\hat{j}'\hat{J}\hat{J}'\left\{ \begin{array}{ccc}
		J & J' & K\\
		j' & j & J_{f}
	\end{array}\right\} \left\{ \begin{array}{ccc}
		J & J' & K\\
		L' & L & S
	\end{array}\right\} \left(L0L'0|K0\right)\;.
\end{align}
This can be simplified by summing over the photon polarizations: 
\begin{align}
	\sum_{\lambda}\lambda^{\kappa+\kappa'}\left(-\right)^{\lambda}\left(j-\lambda j'\lambda|K0\right)= & \left(-\right)^{1}\left(j-1j'1|K0\right)+\left(-\right)^{\kappa+\kappa'-1}\left(j1j'-1|K0\right)\nonumber \\
	= & -\left[1+\left(-\right)^{\kappa+\kappa'+j+j'-K}\right]\left(j-1j'1|K0\right)\nonumber \\
	= & \left(j-1j'1|K0\right) \times \begin{cases}
		-2 & \text{if }(-)^{\kappa+\kappa'+j+j'}=(-)^{K}\;, \\
		0 & \text{otherwise}\;.
	\end{cases}
\end{align}
The result is
\begin{align}
	\hat{s}_{T}^{2}\hat{s}_{P}^{2}\bar{\sum_{i}}\sum_{f}\sum_{\lambda}\left|H_{\lambda}^{fi}\right|^{2}= & \frac{16\pi^{2}}{k^{2}}\sum_{K}P_{K}\sum_{\kappa\kappa'jj'}'\sum_{SLL'JJ'}i^{L-L'+j-j'}\hat{j}\hat{j}'\hat{J}\hat{J}'\hat{L}\hat{L}'e^{i\left(\sigma_{L}-\sigma_{L'}\right)}\nonumber \\
	& \times \left(-\right)^{S-J_{f}+1-L+L'-J+J'}\mathcal{T}_{J_{f}SLJ}^{\kappa j}\mathcal{T}_{J_{f}SL'J'}^{\kappa'j'*}\left\{ \begin{array}{ccc}
		J & J' & K\\
		j' & j & J_{f}
	\end{array}\right\} \left\{ \begin{array}{ccc}
		J & J' & K\\
		L' & L & S
	\end{array}\right\} \nonumber \\
	& \times \left(L0L'0|K0\right)\left(j-1j'1|K0\right)\;,
\end{align}
where $\sum_{\kappa\kappa'jj'}'$ has the condition $(-)^{\kappa+\kappa'+j+j'}=(-)^{K}$.

The angular momentum algebra and matrix elements can be collected into coefficients for the Legendre polynomials, i.e.
\begin{align}
	a_{K} & =\sum_{\kappa\kappa'jj}'\sum_{SLL'JJ'}\left(-\right)^{S-J_{f}+1-L+L'-J+J'}\hat{j}\hat{j}'\hat{J}\hat{J}'\hat{L}\hat{L}'e^{i\left(\sigma_{L}-\sigma_{L'}\right)}i^{L-L'+j-j'}\mathcal{T}_{J_{f}SLJ}^{\kappa j}\mathcal{T}_{J_{f}SL'J'}^{\kappa'j'*}\nonumber \\
	& \times \left\{ \begin{array}{ccc}
		J & J' & K\\
		j' & j & J_{f}
	\end{array}\right\} \left\{ \begin{array}{ccc}
		J & J' & K\\
		L' & L & S
	\end{array}\right\} \left(L0L'0|K0\right)\left(j-1j'1|K0\right)\;.
\end{align}
Inserting everything into (\ref{eq:radcap_ds_Hl}), 
we first integrate
over $\phi_{q}$ to get a factor of $2\pi$ and then move $\dd\cos\theta_{q}$
to the left-hand side to get the differential cross section
\begin{align}
	\frac{\dd\sigma}{\dd\cos\theta_q}= & \frac{4\pi\alpha}{v}\frac{2\pi \omega^{2}\dd \omega}{(2\pi)^{3}2\omega}2\pi\delta(E_{I}-E_{F}-\omega)\bar{\sum_{i}}\sum_{f}\sum_{\lambda}\left|H_{\lambda}^{fi}\right|^{2}\nonumber \\
	= & \frac{\alpha \omega f_{r}}{v\hat{s}_{T}^{2}\hat{s}_{P}^{2}}\frac{16\pi^{2}}{k^{2}}\sum_{K}a_{K}P_{K}(\cos\theta_q)\;,
\end{align}
where $f_{r}=(1+\frac{\omega-p_{P}\cos\theta_q}{M_{F}})^{-1}$ is the recoil
factor, with $p_P$ the projectile momentum and $M_F$ is the mass of the final state bound nucleus. All the $P_{K}(\cos\theta_q)$ with $K\neq0$ integrate to 0
(neglecting recoil).

The integrated cross section is then
\begin{align}
	\sigma= & \frac{32\pi^{2}\alpha \omega}{vk^{2}\hat{s}_{T}^{2}\hat{s}_{P}^{2}}a_{0}\;,
\end{align}
where
\begin{align}
	a_{0}= & \sum_{\kappa\kappa'jj}'\sum_{SLL'JJ'}\left(-\right)^{S-J_{f}+1-L+L'-J+J'}\hat{j}\hat{j}'\hat{J}\hat{J}'\hat{L}\hat{L}'e^{i\left(\sigma_{L}-\sigma_{L'}\right)}i^{L-L'+j-j'}\mathcal{T}_{J_{f}SLJ}^{\kappa j}\mathcal{T}_{J_{f}SL'J'}^{\kappa'j'*}\nonumber \\
	& \times \left\{ \begin{array}{ccc}
		J & J' & 0\\
		j' & j & J_{f}
	\end{array}\right\} \left\{ \begin{array}{ccc}
		J & J' & 0\\
		L' & L & S
	\end{array}\right\} \left(L0L'0|00\right)\left(j-1j'1|00\right)\nonumber \\
	= & \sum_{\kappa j}\sum_{SLJ}\left|\mathcal{T}_{J_{f}SLJ}^{\kappa j}\right|^{2}\;.
\end{align}
%
%
In the long-wavelength limit this becomes
\begin{align}
	\sigma= & \frac{32\pi^{2}\alpha}{vk^{2}\hat{s}_{T}^{2}\hat{s}_{P}^{2}}\sum_{\kappa j}\frac{\omega^{2j+1}}{[(2j+1)!!]^{2}}\frac{j+1}{j}\sum_{L_{i}S_{i}J_{i}}\left|\bra{\Psi^{J_{f}^{\pi_{f}}}}\left|\mathcal{M}_{j}^{\kappa}\right|\ket{\Psi_{\nu_i S_{i}L_{i}}^{J_{i}^{\pi_{i}}}}\right|^{2}\label{eq:radcap_cross}\;,
\end{align}
matching the literature up to normalization conventions \cite{Descouvemont,ThompsonNunes}.

\section{Extended Derivation - Pair Production in Radiative Capture}
\label{sec:pairprod_radcap_appendix}

When pair production occurs during a radiative capture reaction, i.e. $T(P,e^{+}e^{-})F$, as shown in \autoref{fig:feyn_cap_pair}, we must calculate the cross section for a $2\to3$ process. The cross section is
\begin{align}
	\dd\sigma= & \frac{1}{v}\frac{1}{4E_{P}E_{T}}\bar{\sum_{i}}\sum_{f}\sum_{s_{+}s_{-}}\left|\mathcal{M}_{FI}^{s_{+}s_{-}}\right|^{2}\frac{\dd^{3}p_{+}}{(2\pi)^{3}2E_{+}}\frac{\dd^{3}p_{-}}{(2\pi)^{3}2E_{-}}\frac{\dd^{3}p_{F}}{(2\pi)^{3}2E_{F}}\;,
\end{align}
where, as in \autoref{sec:pairprod}, $P_{+}=(E_{+},\vec{p}_{+})$ and $P_{-}=(E_{-},\vec{p}_{-})$
are respectfully the 4-momenta of the emitted positron and electron
and $P_{F}=(E_F,\vec{p}_{F})$ is the 4-momentum of the final state
nucleus. The intermediate photon has 4-momentum $Q=(\omega,\vec{q})$.

The amplitude is evaluated according to the Feynman rules of QED.
We insert the Dirac spinors $\bar{u}^{s_{-}}$ for the outgoing electron
and $\nu^{s_{+}}$ for the outgoing positron, then insert the propagator
$\frac{-i\eta_{\mu\nu}}{Q^{2}}$ for the exchanged photon between
the vertex factors $ie\gamma_{\mu}$ and $ie\mathcal{J}_{\nu}$ where
we use short-hand for the nuclear current $\mathcal{J}_{\nu}=\bra{f}\mathcal{J}_{\nu}(q)\ket{i}$.

The squared amplitude is
\begin{align}
	\left|\mathcal{M}^{s_{+}s_{-}}_{FI}\right|^2=&\left|\bar{u}^{s_{-}}(P_{-})\left(ie\gamma_{\mu}\right)\left(\frac{-i\eta^{\mu\nu}}{Q^{2}}\right)\left(ie\mathcal{J}_{\nu}\right)v^{s_{+}}(P_{+})\right|^2 \\ 
	&\times (2\pi)^{4}\delta(P_{P}+P_T-P_{F}-P_+-P_{-}) 8E_FE_PE_T
	\;.
\end{align}
Here we've integrated over the intermediate 4-momentum $Q$, i.e.
\begin{align}
	& \int\frac{\dd^{4}Q}{(2\pi)^{4}}\frac{1}{Q^4}(2\pi)^{4}\delta(P_P+P_T-P_{F}-Q)(2\pi)^{4}\delta(Q-P_+-P_-)\nonumber \\
	= & \frac{1}{Q^4}(2\pi)^{4}\delta(P_P+P_T-P_F-P_+-P_{-})\;.
\end{align}
In the second line and hereafter the symbols $Q=(\omega,\vec{q})$ are defined by the external momenta:
\begin{align}
	Q= & P_{+}+P_{-}\;,\\
	(\omega,\vec{q})=& (E_++E_-,\vec{p}_++\vec{p}_-)\;.
\end{align}
To simplify the expressions we define $x=\cos\theta_{ee}=\frac{\vec{p}_{+}\cdot\vec{p}_{-}}{|p_+||p_-|}$
and use $m=m_{e}$, e.g.
\begin{align}
	Q^{2}= & 2\left(E_{-}E_{+}-p_{-}p_{+}x+m^{2}\right)\;,\\
	q^{2}= & p_{-}^{2}+p_{+}^{2}+2p_{-}p_{+}x\;.
\end{align}
The lepton tensor is computed by via the lepton currents, summing
over the final state spinor indices, i.e.
\begin{align}
	\ell_{\mu\nu}= & \sum_{s_+s_-} \ell_\mu \ell_\nu^\dagger \\
	=& \sum_{s_+s_-}\left(\bar{u}^{s_{-}}(P_-)\gamma_{\mu}v^{s_{+}}(P_+)\right)\left(\bar{v}^{s+}(P_+)\gamma_{\nu}u^{s_{-}}(P_-)\right)\nonumber \\
	= & \text{Tr}\left[\gamma_{\mu}\left(\slashed{P}_{-}+m\right)\gamma_{\nu}\left(\slashed{P}_{+}-m\right)\right]\nonumber \\
	= & 4\left[P_{-\mu}P_{+\nu}+P_{-\nu}P_{+\mu}-\eta_{\mu\nu}\left(P_{-}\cdot P_{+}+m^{2}\right)\right]\label{eq:lepton_tensor}\;.
\end{align}
Folding this with the nuclear current, we have
\begin{align}
	\sum_{s_{+}s_{-}}\left|\bar{u}^{s_{-}}(P_{-})\left(ie\gamma_{\mu}\right)\left(\frac{-i\eta^{\mu\nu}}{Q^{2}}\right)\left(ie\mathcal{J}_{\nu}\right)v^{s_{+}}(P_{+})\right|^2= & \frac{e^4}{Q^4}\ell^{\mu\nu}\mathcal{J}_{\mu}^{*}\mathcal{J}_{\nu}\;.
\end{align}
The resulting differential cross section is
\begin{align}
	\dd\sigma= & \frac{1}{v}\frac{1}{4E_{P}E_{T}}\frac{\dd^{3}p_{+}}{(2\pi)^{3}2E_{+}}\frac{\dd^{3}p_{-}}{(2\pi)^{3}2E_{-}}\frac{\dd^{3}p_{F}}{(2\pi)^{3}2E_{F}}8E_{P}E_{T}E_{F}\nonumber \\
	&\times (2\pi)^{4}\delta^{4}(P_{P}+P_{T}-P_{F}-P_{+}-P_{-}) \frac{e^{4}}{Q^{4}}\bar{\sum_{i}}\sum_{f}\ell^{\mu\nu}\mathcal{J}_{\mu}\mathcal{J}_{\nu}^{*}\nonumber \\
	= & \frac{1}{v}\frac{e^{4}}{Q^{4}}\frac{\dd^{3}p_{+}}{(2\pi)^{3}2E_{+}}\frac{\dd^{3}p_{-}}{(2\pi)^{3}2E_{-}}2\pi\delta(E_{P}+E_{T}-E_{F}-E_{+}-E_{-})\bar{\sum_{i}}\sum_{f}\ell^{\mu\nu}\mathcal{J}_{\mu}\mathcal{J}_{\nu}^{*}\;.\label{eq:intermediate_dsigma}
\end{align}
by integration over the final nuclear momentum.

The contraction between the lepton tensor and nuclear currents mixes the nuclear operators with different character, resulting in a cross section with six different terms. Following the notation of \cite{Viviani2022}, we have
\begin{align}
	\frac{1}{4}\bar{\sum_{i}}\sum_{f}\ell^{\mu\nu}\mathcal{J}_{\mu}\mathcal{J}_{\nu}^{*}= & \sum_{n=1}^{6}v_{n}R_{n}\;.
\end{align}
The expansion of the contraction is done by collecting like terms,
here we make use of $\mathcal{J}=(\rho,\mathcal{J}_{x},\mathcal{J}_{y},\mathcal{J}_{z})$
explicitly where $\mathcal{J}_{z}=\frac{\omega}{\left|\vec{q}\right|}\rho$.
The lepton-nuclear tensor contraction is then
\begin{align}
	\frac{1}{4}\ell^{\mu\nu}\mathcal{J}_{\mu}\mathcal{J}_{\nu}^{*}= & \left(P_{+}\cdot\mathcal{J}\right)\left(P_{-}\cdot\mathcal{J}^{*}\right)+\left(P_{+}\cdot\mathcal{J}^{*}\right)\left(P_{-}\cdot\mathcal{J}\right)-\left(\mathcal{J}\cdot\mathcal{J}^{*}\right)\left(P_{+}\cdot P_{-}+m^{2}\right)\;,
\end{align}
where the ``$\cdot$'' operator here represents a contraction of four-vectors, e.g. $\mathcal{J} \cdot \mathcal{J}^* = \mathcal{J}^\mu \mathcal{J}_\mu^*=\rho^2-\vec{\mathcal{J}}\cdot\vec{\mathcal{J}}^*$.

Let's define
$A=P_{+}$ and $B=P_{-}$ to avoid typing so many subscripts. $Q=A+B$
and we define $C=A-B$ (such that $A=\frac{1}{2}(C+Q)$, $B=\frac{-1}{2}(C-Q)$).
This simplifies the equation, i.e.
\begin{align}
	\frac{1}{4}\ell^{\mu\nu}\mathcal{J}_{\mu}\mathcal{J}_{\nu}^{*}= & -\left|\mathcal{J}\right|^{2}\left(A\cdot B+m^{2}\right)-\frac{1}{2}\left|C\cdot\mathcal{J}\right|^{2}\;.
\end{align}
%
We write $A=(\alpha,\vec{a})=(E_{+},\vec{p}_+)$ and $B=(\beta,\vec{b})=(E_{-},\vec{p}_-)$ so $\omega=\alpha+\beta$ and $\vec{q}=\vec{a}+\vec{b}$. Finally, $C=(\varsigma,\vec{c})=(\varsigma,c_{x},c_{y},c_{z})$ where $\varsigma=\alpha-\beta$ and $\vec{c}=\vec{a}-\vec{b}$.
Expanding, we have
\begin{align}
	\frac{1}{4}\ell^{\mu\nu}\mathcal{J}_{\mu}\mathcal{J}_{\nu}^{*}= & =-\left(\rho^{2}-\mathcal{J}_{x}^{2}-\mathcal{J}_{y}^{2}-\mathcal{J}_{z}^{2}\right)\left(\alpha\beta - \vec{a}\cdot \vec{b}+m^{2}\right)-\frac{1}{2}\bigg[\varsigma^{2}\rho^{2}+c_{x}^{2}\mathcal{J}_{x}^{2}+c_{y}^{2}\mathcal{J}_{y}^{2}+c_{z}^{2}\mathcal{J}_{z}^{2}\nonumber \\
	& -\varsigma c_{x}\left(\rho \mathcal{J}_{x}^{*}+\mathcal{J}_{x}\rho^{*}\right)-\varsigma c_{y}\left(\rho \mathcal{J}_{y}^{*}+\mathcal{J}_{y}\rho^{*}\right)-\varsigma c_{z}\left(\rho \mathcal{J}_{z}^{*}+\mathcal{J}_{z}\rho^{*}\right)\nonumber \\
	& +c_{x}c_{y}\left(\mathcal{J}_{x}\mathcal{J}_{y}^{*}+\mathcal{J}_{y}\mathcal{J}_{x}^{*}\right)+c_{x}c_{z}\left(\mathcal{J}_{x}\mathcal{J}_{z}^{*}+\mathcal{J}_{z}\mathcal{J}_{x}^{*}\right)+c_{y}c_{z}\left(\mathcal{J}_{y}\mathcal{J}_{z}^{*}+\mathcal{J}_{z}\mathcal{J}_{y}^{*}\right)\bigg]\;.
\end{align}

\subsection{Collection of Like Terms}
We collect like terms (using the shorthand $\rho^2=\rho \rho^*$ and suppressing the sums $\bar{\sum}_i \sum_f$).

\paragraph{Term 1: $\rho^{2}$, $\mathcal{J}_{z}^{2}$, $\rho \mathcal{J}_{z}^{*}$, $\mathcal{J}_{z}\rho^{*}$}

\begin{align}
	v_{1}R_{1}= & -\left(\rho^{2}-\mathcal{J}_{z}^{2}\right)\left(\alpha\beta-\vec{a}\cdot\vec{b}+m^{2}\right)-\frac{1}{2}\left(\varsigma^{2}\rho^{2}-\varsigma c_{z}\rho \mathcal{J}_{z}^{*}-c_{z}\varsigma \mathcal{J}_{z}\rho^{*}+c_{z}^{2}\mathcal{J}_{z}^{2}\right)\nonumber \\
	= & -\rho^{2}\left(1-\frac{\omega^{2}}{q^{2}}\right)\left(\alpha\beta-\vec{a}\cdot\vec{b}+m^{2}\right)-\frac{1}{2}\left(\varsigma-c_{z}\frac{\omega}{\left|\vec{q}\right|}\right)^{2}\rho^{2}\label{eq:v1R1}\;.
\end{align}
Simplification requires some manipulation. We use $c_{z}=\frac{\vec{c}\cdot\vec{q}}{\left|\vec{q}\right|}$,
and some identities:
\begin{align}
	\vec{c}\cdot\vec{q}= & b^{2}-a^{2}\nonumber \\
	= & \beta^{2}-\alpha^{2}\nonumber \\
	= & \varsigma \omega\;,
\end{align}
\begin{align}
	Q^{2}= & \omega^{2}-q^{2}\;,\\
	\frac{Q^{2}}{2}= & \alpha\beta-\vec{a}\cdot\vec{b}+m^{2}\;.
\end{align}

We take $R_{1}\equiv\rho^{2}$ and so
\begin{align}
	v_{1}= & \left(\frac{Q^{2}}{q^{2}}\right)\left(\alpha\beta-\vec{a}\cdot\vec{b}+m^{2}\right)-\frac{1}{2}\left(\varsigma-\frac{\varsigma\omega}{\left|\vec{q}\right|}\frac{\omega}{\left|\vec{q}\right|}\right)^{2}\nonumber \\
	= & \left(\frac{Q^2}{q^2}\right)\left(\frac{Q^2}{2}\right) -\frac{\varsigma^2}{2}\left(-\frac{Q^2}{q^2}\right)^2 \nonumber \\
	= & \left(\frac{Q^4}{q^4}\right)\left(\frac{q^2 -\varsigma^2}{2}\right) \nonumber \\
	= & \left(\frac{Q^{4}}{q^{4}}\right)\left(\alpha\beta+\vec{a}\cdot\vec{b}-m^{2}\right)\label{eq:v1}\;.
\end{align}

\paragraph{Term 2: $\rho \mathcal{J}_{x}^{*}$, $\mathcal{J}_{x}\rho^{*}$, $\mathcal{J}_{z}\mathcal{J}_{x}^{*}$, $\mathcal{J}_{x}\mathcal{J}_{z}^{*}$}

\begin{align}
	v_{2}R_{2}= & -\frac{1}{2}\left[-\varsigma c_{x}\left(\rho \mathcal{J}_{x}^{*}+\mathcal{J}_{x}\rho^{*}\right)+c_{z}c_{x}\left(\mathcal{J}_{z}\mathcal{J}_{x}^{*}+\mathcal{J}_{x}\mathcal{J}_{z}^{*}\right)\right]\nonumber \;,\\
	v_{2}\equiv & -\frac{1}{\sqrt{2}}c_{x}\left(\alpha-\beta-\frac{\omega}{\left|\vec{q}\right|}c_{z}\right)\;,\\
	R_{2}\equiv & \text{Re}\left(\rho^{*}\left(\mathcal{J}_{+}-\mathcal{J}_{-}\right)\right)\;,
\end{align}
where we've used:
\begin{align}
	\mathcal{J}_{x}\rho^{*}+\rho \mathcal{J}_{x}^{*}= & 2\text{Re}(\mathcal{J}_{x}\rho^{*})\nonumber \\
	= & 2\text{Re}(\frac{-1}{\sqrt{2}}\left(\mathcal{J}_{+}-\mathcal{J}_{-}\right)\rho^{*})\;.
\end{align}

\paragraph{Term 3: $\rho \mathcal{J}_{y}^{*}$, $\mathcal{J}_{y}\rho^{*}$, $\mathcal{J}_{z}\mathcal{J}_{y}^{*}$, $\mathcal{J}_{y}\mathcal{J}_{z}^{*}$}

\begin{align}
	v_{3}R_{3}= & -\frac{1}{2}\left[-\varsigma c_{y}\left(\rho \mathcal{J}_{y}^{*}+\mathcal{J}_{y}\rho^{*}\right)+c_{z}c_{y}\left(\mathcal{J}_{z}\mathcal{J}_{y}^{*}+\mathcal{J}_{y}\mathcal{J}_{z}^{*}\right)\right]\nonumber \;,\\
	v_{3}\equiv & -\frac{1}{\sqrt{2}}c_{y}\left(\alpha-\beta-\frac{\omega}{\left|\vec{q}\right|}c_{z}\right)\;,\\
	R_{3}\equiv & \text{Im}\left(\rho^{*}\left(\mathcal{J}_{+}+\mathcal{J}_{-}\right)\right)\;.
\end{align}
Since $\mathcal{J}_{y}$ has the additional factor of $i$:
\begin{align}
	\mathcal{J}_{y}\rho^{*}+\rho \mathcal{J}_{y}^{*}= & 2\text{Re}(\mathcal{J}_{y}\rho^{*})\nonumber \\
	= & 2\text{Im}(\frac{-1}{\sqrt{2}}\left(\mathcal{J}_{+}+\mathcal{J}_{-}\right)\rho^{*})\;.
\end{align}

\paragraph{$\mathcal{J}_{x},\mathcal{J}_{y}$ Cross Terms}:\\

The remaining terms are
\begin{equation}
	\left(\left|\mathcal{J}_{x}\right|^{2}+\left|\mathcal{J}_{y}\right|^{2}\right)\left(A\cdot B+m^{2}\right)-\frac{1}{2}\left(c_{x}^{2}\left|\mathcal{J}_{x}\right|^{2}+c_{y}^{2}\left|\mathcal{J}_{y}\right|^{2}+c_{x}c_{y}\left(\mathcal{J}_{x}\mathcal{J}_{y}^{*}+\mathcal{J}_{y}\mathcal{J}_{x}^{*}\right)\right)\;.
\end{equation}
We use
\begin{align}
	\left|\mathcal{J}_x\right|^2 = & \frac{1}{2}\left(\left|\mathcal{J}_+\right|^2 - \mathcal{J}_+\mathcal{J}_-^* - \mathcal{J}_-\mathcal{J}_+^* + \left|\mathcal{J}_-\right|^2\right) \;, \\
	\left|\mathcal{J}_y\right|^2 = & \frac{1}{2}\left(\left|\mathcal{J}_+\right|^2 + \mathcal{J}_+\mathcal{J}_-^* + \mathcal{J}_-\mathcal{J}_+^* + \left|\mathcal{J}_-\right|^2\right) \;, \\
	\mathcal{J}_x \mathcal{J}_y^* = & \frac{-i}{2}\left(\left|\mathcal{J}_+\right|^2 + \mathcal{J}_+\mathcal{J}_-^* - \mathcal{J}_-\mathcal{J}_+^* - \left|\mathcal{J}_-\right|^2\right) \;, \\
	\mathcal{J}_y \mathcal{J}_x^* = & \frac{i}{2}\left(\left|\mathcal{J}_+\right|^2 - \mathcal{J}_+\mathcal{J}_-^* + \mathcal{J}_-\mathcal{J}_+^* - \left|\mathcal{J}_-\right|^2\right) \;,
\end{align}
and
\begin{align}
	\left|\mathcal{J}_{x}\right|^{2}+\left|\mathcal{J}_{y}\right|^{2}= & \left|\mathcal{J}_{+}\right|^{2}+\left|\mathcal{J}_{-}\right|^{2} \;, \\
	\mathcal{J}_{x}\mathcal{J}_{y}^{*}+\mathcal{J}_{y}\mathcal{J}_{x}^{*}= & -i\left(\mathcal{J}_{+}\mathcal{J}_{-}^{*}-\mathcal{J}_{-}\mathcal{J}_{+}^{*}\right)\;.
\end{align}
Then
\begin{align}
	& c_{x}^{2}\left|\mathcal{J}_{x}\right|^{2}+c_{y}^{2}\left|\mathcal{J}_{y}\right|^{2}+c_{x}c_{y}\left(\mathcal{J}_{x}\mathcal{J}_{y}^{*}+\mathcal{J}_{y}\mathcal{J}_{x}^{*}\right)\nonumber \\
	= & \frac{1}{2}\left(c_{x}^{2}+c_{y}^{2}\right)\left(\left|\mathcal{J}_{+}\right|^{2}+\left|\mathcal{J}_{-}\right|^{2}\right)\nonumber \\
	& -\frac{1}{2}\left(c_{x}^{2}-c_{y}^{2}\right)\left(\mathcal{J}_{+}\mathcal{J}_{-}^{*}+\mathcal{J}_{-}\mathcal{J}_{+}^{*}\right)\nonumber \\
	& -ic_{x}c_{y}\left(\mathcal{J}_{+}\mathcal{J}_{-}^{*}-\mathcal{J}_{-}\mathcal{J}_{+}^{*}\right)\;.
\end{align}

\paragraph{Term 4: $\left|\mathcal{J}_{+}\right|^{2}$, $\left|\mathcal{J}_{-}\right|^{2}$}

\begin{align}
	v_{4}R_{4}= & \left(\left|\mathcal{J}_{+}\right|^{2}+\left|\mathcal{J}_{-}\right|^{2}\right)\left(\alpha\beta-\vec{a}\cdot\vec{b}+m^{2}\right)-\frac{1}{2}\cdot\frac{1}{2}\left(c_{x}^{2}+c_{y}^{2}\right)\left(\left|\mathcal{J}_{+}\right|^{2}+\left|\mathcal{J}_{-}\right|^{2}\right)\nonumber \\
	= & \left[-\frac{1}{4}\left(c_{x}^{2}+c_{y}^{2}\right)+\alpha\beta-\vec{a}\cdot\vec{b}+m^{2}\right]\left[\left|\mathcal{J}_{+}\right|^{2}+\left|\mathcal{J}_{-}\right|^{2}\right]\nonumber\;, \\
	v_{4}\equiv & -\frac{1}{4}\left(c_{x}^{2}+c_{y}^{2}\right)+\alpha\beta-\vec{a}\cdot\vec{b}+m^{2}\;,\\
	R_{4}\equiv & \left|\mathcal{J}_{+}\right|^{2}+\left|\mathcal{J}_{-}\right|^{2}\;.
\end{align}

\paragraph{Term 5: $\mathcal{J}_{+}\mathcal{J}_{-}^{*}+\mathcal{J}_{-}\mathcal{J}_{+}^{*}$}

\begin{align}
  v_{5}R_{5} = &
         -\frac{1}{2}\left[-\frac{1}{2}(c^2_{x}-c^2_{y})\left(\mathcal{J}_{+}\mathcal{J}_{-}^{*}-\mathcal{J}_{-}\mathcal{J}_{+}^{*}\right)\right]\nonumber \;\\
        = & \frac{1}{2}\left(c_{x}^{2}-c_{y}^{2}\right)\text{Re}\left(\mathcal{J}_{+}^{*}\mathcal{J}_{-}\right)\nonumber\;, \\
	v_{5}\equiv & \frac{1}{2}\left(c_{x}^{2}-c_{y}^{2}\right)\;,\\
	R_{5}\equiv & \text{Re}\left(\mathcal{J}_{+}^{*}\mathcal{J}_{-}\right)\;.
\end{align}

\paragraph{Term 6: $\mathcal{J}_{+}\mathcal{J}_{-}^{*}-\mathcal{J}_{-}\mathcal{J}_{+}^{*}$}

\begin{align}
	v_{6}R_{6}= & -\frac{1}{2}\left[-ic_{x}c_{y}\left(\mathcal{J}_{+}\mathcal{J}_{-}^{*}-\mathcal{J}_{-}\mathcal{J}_{+}^{*}\right)\right]\nonumber \;\\
	= & -c_{x}c_{y}\text{Im}\left(\mathcal{J}_{+}^{*}\mathcal{J}_{-}\right)\nonumber\;, \\
	v_{6}\equiv & -c_{x}c_{y}\\
	R_{6}\equiv & \text{Im}\left(\mathcal{J}_{+}^{*}\mathcal{J}_{-}\right)\;.
\end{align}

\begin{table}[tbph]
	\centering
	\begin{tabular}{|c|c|c|c|}
		\hline
		$n$ & $v_n$ & $R_n$ & $N_{\mu}N_{\nu}^{*}$\\
		\hline
		\hline
		$1$ & $\frac{Q^{4}}{q^{4}}(\alpha\beta+\vec{a}\cdot\vec{b}-m^{2})$ & $\rho^{2}$ & $N_{0}N_{0}^{*}$\\ \hline
		$2$ & $-\frac{1}{\sqrt{2}}c_x\left(\alpha-\beta-\frac{\omega}{q}c_z\right)$  & $\text{Re}(\rho(\mathcal{J}_{+}-\mathcal{J}_{-})^{*}$) & $\text{Re}(N_{0}(N_{+}-N_{-})^{*})$\\ \hline
		$3$ & $-\frac{1}{\sqrt{2}}c_y\left(\alpha-\beta-\frac{\omega}{q}c_z\right)$ & $-\text{Im}(\rho(\mathcal{J}_{+}+\mathcal{J}_{-})^{*})$ & $-\text{Im}(N_{0}(N_{+}+N_{-})^{*})$\\ \hline
		$4$ & $-\frac{1}{4}(c_x^2+c_y^2)+\alpha\beta-\vec{a}\cdot\vec{b}+m^2$ & $\mathcal{J}_{+}^{2}+\mathcal{J}_{-}^{2}$ & $N_{+}N_{+}^{*}+N_{-}N_{-}^{*}$\\ \hline
		$5$ & $\frac{1}{2}\left(c_x^2-c_y^2\right)$  & $\text{Re}(\mathcal{J}_{+}\mathcal{J}_{-}^{*})$ & $\text{Re}(N_{+}N_{-}^{*})$\\ \hline
		$6$ & $-c_xc_y$ & $-\text{Im}(\mathcal{J}_{+}\mathcal{J}_{-}^{*})$ & $-\text{Im}(N_{+}N_{-}^{*})$\\
		\hline
	\end{tabular}
	\caption{Summary of the kinematic prefactors $v_n$ and corresponding nuclear transition operators $R_n$. They are in agreement with Ref. \cite{Viviani2022}. Each $R_n$ corresponds to an $N_\mu N_\nu^*$ combination, as defined in (\ref{eq:NN}).}
	\label{tab:v_R_NN}
\end{table}

\subsection{Differential Cross Section}

The six terms will be inserted in to the differential cross section (\ref{eq:intermediate_dsigma}) after shifting
to spherical coordinates $\dd^{3}p_\pm=p_\pm^{2}\dd p_\pm\dd\Omega_\pm$, changing variables $p_{+}\dd p_{+}=E_{+}\dd E_{+}$ and using
the $\delta$ to do the integral over $E_{-}$, i.e.
\begin{align}
	\dd\sigma= & \frac{4}{v}\frac{\alpha^{2}}{Q^{4}}f_{r}\frac{p_{-}p_{+}\dd E_{+}\dd\Omega_{+}\dd\Omega_{-}}{(2\pi)^{3}}\sum_{n}v_{n}R_{n}\nonumber \;,\\
	\frac{\dd^{5}\sigma}{\dd E_{+}\dd\Omega_{+}\dd\Omega_{-}}= & \frac{4\alpha^{2}}{(2\pi)^{3}}\frac{f_{r}}{v}\frac{p_{+}p_{-}}{Q^{4}}\sum_{n}v_{n}R_{n}\label{eq:full_diff_sigma}\;,
\end{align}
where appears the recoil factor $f_{r}=\left(1+\left(\frac{E_{-}}{p_{-}}\right)\frac{p_{-}+p_{+}x-p_{P}\cos\theta_{-}}{E_{F}}\right)^{-1}$ (which is negligible ($f_r\simeq1$) in the energy range considered in our calculations).

For radiative capture we have to compute on the matrix elements of
the transverse electromagnetic current (i.e. with $\lambda=\pm1$).
Again those operators are $X_{j-\lambda}^{\kappa}=\frac{\sqrt{4\pi}}{k}\sqrt{2\pi}\hat{j}(-i)^{j}\mathcal{T}_{j-\lambda}^{\kappa}$
where $\kappa$ denotes electric or magnetic multipoles ($\kappa=0,1$
respectively). In pair production we have a virtual photon and hence
the polarization can also be $\lambda=0$ (longitudinal). This is apparent due to
the presence of $\bra{f}\rho\ket{i}$ terms in the cross section.

To evaluate both $\rho$ and $\mathcal{J}$ terms (and their combinations)
we define the operator $\mathcal{N}_{jm}^{\mu}$ where $\mu=0$
corresponds to the charge operator $\rho$. $\mu=\pm1$ corresponds
to the current operator with $\lambda=\pm1$: $\mathcal{N}_{jm}^{\lambda}=\sum_{\kappa} \lambda^\kappa X_{jm}^{\kappa}$
($H_{\lambda}$ in \autoref{sec:radcap_appendix}). We have
\begin{align}
	\mathcal{N}_{jm}^{0}=&\frac{\sqrt{4\pi}}{k}\sqrt{4\pi}\hat{j}(-i)^{j}\mathcal{C}_{jm} \;,\\
	\mathcal{N}_{jm}^{\lambda}=&\frac{\sqrt{4\pi}}{k}\sqrt{2\pi}\hat{j}(-i)^{j}\sum_{\kappa}\lambda^{\kappa}\mathcal{T}_{jm}^{\kappa}\;.
\end{align}

We will have to multiply
by a more general $D$-matrix $D_{m\mu}^{j}$ where $\mu\in(0,\pm1)$. If we define the leptonic coordinate system to have an axis perpendicular to both the internal photon and initial velocity, i.e. $\hat{z}=\hat{q}$ and $\hat{y} = \hat{v}\times\hat{q}$ ($\hat{x} = \hat{y}\times\hat{z}$), the necessary rotation is $\mathcal{R}(\phi_q,\theta_q,0)$ and so $D^j_{m-\mu}(-\phi_q,-\theta_q,0)$ appears.
%
%
%

Each $R_{n}$ term corresponds to one of the combinations of transition matrix elements listed in Table \ref{tab:v_R_NN}, where
\begin{equation}
	N_{\mu}N_{\nu}^{*}=\bar{\sum_i}\sum_{f}\bra{f}\sum_{jm}D_{m-\mu}^{j}\mathcal{N}_{jm}^{\mu}\ket{i}\left(\bra{f}\sum_{j'm'}D_{m'-\nu}^{j'}\mathcal{N}_{j'm'}^{\nu}\ket{i}\right)^{*}\;.
\end{equation}

A general expression for $N_{\mu}N_{\nu}^{*}$ follows from very similar steps to \autoref{sec:radcap_appendix}, i.e.
using the notation
\begin{equation}
	\mathcal{N}_{J_{f}SLJ}^{\mu j}=\bra{J_{f}}|\mathcal{N}_{j}^{\mu}|\ket{SLJ}\;,
\end{equation}
analogous to $X_{J_{f}SLJ}^{\kappa j}$,
and the property $D^K_{0M}(-\phi_q,-\theta_q,0) = d^K_{0M}(-\theta_q)=d^K_{M0}(\theta_q)$, we have
\begin{align}
	N_{\mu}N_{\nu}^{*}= & \bar{\sum_{m_{T}m_{P}}}\sum_{M_{f}}\left(\bra{J_{f}M_{f}}\sum_{jm}D_{m-\mu}^{j}\mathcal{N}_{jm}^{\mu}\ket{s_{T}m_{T}s_{P}m_{P}}\right)\left(\bra{J_{f}M_{f}}\sum_{j'm'}D_{m'-\nu}^{j'}\mathcal{N}_{j'm'}^{\nu}\ket{s_{T}m_{T}s_{P}m_{P}}\right)^{*}\nonumber \\
	= & (\hat{s}_T\hat{s}_P\hat{J}_{f})^{-2}\sum_{M_{f}}\sum_{jm}\sum_{LSJM}\sum_{j'}\sum_{L'J'}i^{L-L'}\hat{L}\hat{L'}e^{i(\sigma_{L}-\sigma_{L'})}D_{m-\mu}^{j}D_{m-\nu}^{j'*}\mathcal{N}_{J_{f}SLJ}^{\mu j}\mathcal{N}_{J_{f}SL'J'}^{\nu j'*}\nonumber \\
	& \times \left(SML0|JM\right)\left(SML'0|J'M\right)\left(JMjm|J_{f}M_{f}\right)\left(J'Mj'm|J_{f}M_{f}\right)\nonumber \\
	= & (\hat{s}_T\hat{s}_P\hat{J}_{f})^{-2}\sum_{M_{f}}\sum_{jm}\sum_{LSJM}\sum_{j'}\sum_{L'J'}i^{L-L'}\hat{L}\hat{L'}e^{i(\sigma_{L}-\sigma_{L'})}\mathcal{N}_{J_{f}SLJ}^{\mu j}\mathcal{N}_{J_{f}SL'J'}^{\nu j'*}\nonumber \\
	& \times \sum_{K}(-)^{m+\nu}\left(jmj'-m|K0\right)\left(j-\mu j'\nu|K(\nu-\mu)\right)D_{0(\nu-\mu)}^{K}(-\phi_q,-\theta_q,0) \nonumber\\
	& \times \left(SML0|JM\right)\left(SML'0|J'M\right)\left(JMjm|J_{f}M_{f}\right)\left(J'Mj'm|J_{f}M_{f}\right)\nonumber \\
	= & 
	\frac{1}{\hat{s}_T^2\hat{s}_P^2}\sum_{jj'}\sum_{SLL'JJ'}\left(-\right)^{S-J_{f}+\nu-j-j'-L-L'-J+J'}\hat{J}\hat{J}'i^{L-L'}\hat{L}\hat{L'}e^{i(\sigma_{L}-\sigma_{L'})}\mathcal{N}_{J_{f}SLJ}^{\mu j}\mathcal{N}_{J_{f}SL'J'}^{\nu j'*}\nonumber \\
	& \times \sum_{K}\left\{ \begin{array}{ccc}
		J & J' & K\\
		j' & j & J_{f}
	\end{array}\right\} \left\{ \begin{array}{ccc}
		J & J' & K\\
		L' & L & S
	\end{array}\right\} \left(L0L'0|K0\right)\left(j-\mu j'\nu|K(\nu-\mu)\right)d_{(\nu-\mu)0}^{K}(\theta_q)\label{eq:NN}\;.
\end{align}

\begin{figure}[tbph]
	\centering
	\includegraphics[width=0.65\textwidth]{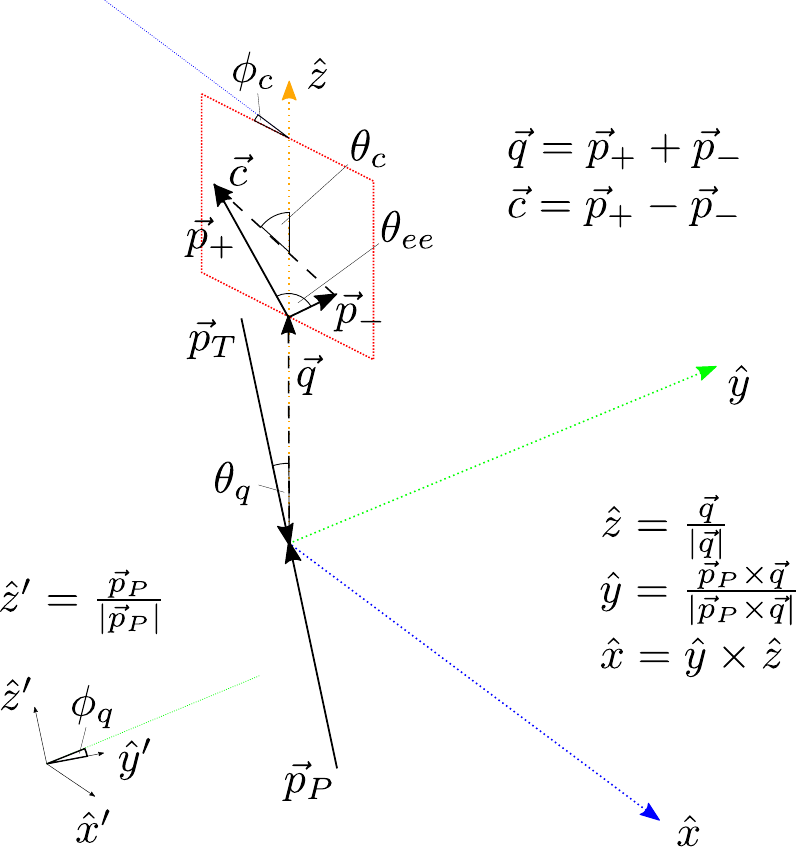}
	\caption{Diagram of relevant vectors and angles.}
	\label{fig:vectors}
\end{figure}

The full differential cross section is difficult to compute explicitly
in the 5-dimensional phase space but comparison can be made to experiment at a fixed set of angles.
To compare to ATOMKI data we set $\phi_q=0$, $\theta_q=\frac{\pi}{2}$ and $\phi_c=\frac{\pi}{2}$ according to the diagram in Fig. \ref{fig:vectors}.
The photon is perpendicular to the beam such that the
electron and positron are in the $y-z$ plane i.e. $c_{z}=\frac{\vec{c}\cdot\vec{q}}{q}$
, $c_{y}=\vec{c}-\frac{\vec{c}\cdot\vec{q}}{q^{2}}\vec{q}$ and $c_{x}=0$.
In this case, $v_{2}=v_{6}=0$,
in addition $R_{3}=0$, and so we actually only need $c_{z}^{2}$
and $c_{y}^{2}$.
Using
\begin{align}
	c_{z}^{2}= & \frac{(\vec{c}\cdot\vec{q})^{2}}{q^{2}}\nonumber \\
	= & \frac{((\vec{a}-\vec{b})\cdot(-\vec{a}-\vec{b}))^{2}}{q^{2}}\nonumber \\
	= & \frac{(b^{2}-a^{2})^{2}}{q^{2}}\;,
\end{align}
and
\begin{align}
	c_{y}^{2}= & \left|\vec{c}-\frac{\vec{c}\cdot\vec{q}}{q^{2}}\vec{q}\right|^{2}\nonumber \\
	= & c^{2}-\frac{(\vec{c}\cdot\vec{q})^{2}}{q^{2}}\nonumber \\
	= & a^{2}-2\vec{a}\cdot\vec{b}+b^{2}-\frac{(b^{2}-a^{2})^{2}}{q^{2}}\;.
\end{align}
The kinematics only depend on the relative angle $\theta_{ee}$, referred to by $\Theta$ in the main text (again $x=\cos\theta_{ee}$).
Table \ref{tab:v} shows the resulting kinematic terms which contribute.
Table \ref{tab:D} shows the simplified expressions that result for $d_{(\mu-\nu)0}^{K}(\theta_q)$ which contribute to the $R_n$
(through Eq. (\ref{eq:NN})).

For a given total energy transfer $\omega=E_++E_-$, the positron and electron divide the energy according to the asymmetry parameter $y$ defined by
\begin{equation}
	y = \frac{E_{+}-E_{-}}{\omega} \label{eq:y}\;.
\end{equation}
The ATOMKI experiment is compared to the differential cross section partially integrated over $y$, i.e.
\begin{align}
	\frac{\dd^4\sigma}{\dd\Omega_+\dd\Omega_-}(\theta_{ee}) = \int \dd y \frac{2\alpha^2}{(2\pi)^3}\frac{\omega f_r}{v}\frac{p_+p_-}{Q^4}\sum_{n\in\{1,4,5\}}v_n R_n \;.\label{eq:partial_sigma}
\end{align}
The domain of $y$ is restricted by the experimental constraints, e.g. $|y|\leq0.5$ in \cite{Krasznahorkay2016} and $|y|\leq0.3$ in \cite{Sas:2022pgm}. The kinematically allowed region is $|y|<\frac{2m}{\omega}$.

\begin{table}[tbph]
	\centering
	\begin{tabular}{|c|c|}
		\hline 
		$n$ &  $v_{n}(\theta_{q}=\frac{\pi}{2},\phi_{c}=\frac{\pi}{2})$\tabularnewline
		\hline 
		\hline 
		1 &  $\frac{Q^{4}}{q^{4}}(\alpha\beta+abx-m^{2})$\tabularnewline
		\hline 
		4 &  $-\frac{1}{4}\left(a^{2}+b^{2}-2abx-\frac{(b^{2}-a^{2})^{2}}{q^{2}}\right)+\alpha\beta-abx+m^{2}$\tabularnewline
		\hline 
		5 &  $-\frac{1}{2}\left(a^{2}+b^{2}-2abx-\frac{(b^{2}-a^{2})^{2}}{q^{2}}\right)$\tabularnewline
		\hline 
	\end{tabular}
	\caption{Simplified formulation of the non-zero kinematic factors $v_n$ corresponding to the ATOMKI experimental setup.}
	\label{tab:v}
\end{table}

\begin{table}[tbph]
	\centering
	\begin{tabular}{|c|c|c|}
		\hline 
		$K,M$ & $d_{M0}^{K}(\theta)$ & $d_{M0}^{K}(\frac{\pi}{2})$\tabularnewline
		\hline 
		\hline 
		$0,0$ & 1 & 1\tabularnewline
		\hline 
		$1,0$ & $\cos\theta$ & 0\tabularnewline
		\hline 
		$1,\pm1$ & $\mp\frac{1}{\sqrt{2}}\sin\theta$ & $\mp\frac{1}{\sqrt{2}}$\tabularnewline
		\hline 
		$2,0$ & $\frac{1}{2}\left(3\cos^{2}\theta-1\right)$ & $-\frac{1}{2}$\tabularnewline
		\hline 
		$2,\pm1$ & $\mp\sqrt{\frac{3}{2}}\sin\theta\cos\theta$ & 0\tabularnewline
		\hline 
		$2,\pm2$ & $\sqrt{\frac{3}{8}}\sin^{2}\theta$ & $\sqrt{\frac{3}{8}}$\tabularnewline
		\hline
  		$3,0$ & $-\frac{1}{2}\cos\theta\left(3-5\cos^{2}\theta\right)$ & $0$\tabularnewline
		\hline 
		$3,\pm1$ & $\pm\frac{\sqrt{3}}{4}\sin\theta\left(1-5\cos^2\theta\right)$ & $\pm\frac{\sqrt{3}}{4}$\tabularnewline
		\hline 
		$3,\pm2$ & $\frac{\sqrt{30}}{4}\sin^2\theta\cos\theta$ & $0$\tabularnewline
		\hline 
  		$4,0$ & $\frac{1}{8}\left(3-30\cos^2\theta+35\cos^4\theta\right)$ & $\frac{3}{8}$\tabularnewline
		\hline 
		$4,\pm1$ & $\pm\frac{\sqrt{5}}{4}\sin\theta\cos\theta\left(3-7\cos^2\theta\right)$ & 0\tabularnewline
		\hline 
		$4,\pm2$ & $-\frac{\sqrt{10}}{8}\sin^2\theta\left(1-7\cos^2\theta\right)$ & $-\frac{\sqrt{10}}{8}$\tabularnewline
		\hline 
	\end{tabular}
	\caption{Relevant values of the Wigner $d$-matrix factor \cite{Varshalovich}.}
	
	\label{tab:D}
\end{table}

\subsection{Integration of the Total Cross Section}\label{sec:pair_inegrated_appendix}

The differential cross section (\ref{eq:full_diff_sigma}) is expressed in terms of 5 variables: $E_+$, $\theta_+$, $\phi_+$, $\theta_-$ and $\phi_-$.
Using a change of variables the integration may be carried out in terms of $y$, $x$, $\theta_q$, $\phi_q$ and $\phi_c$ defined in \autoref{fig:vectors}.
$\theta_q$ and $\phi_q$ are the angles which define the direction of $\vec{q}$ relative to coordinates defined by the initial proton velocity.
$\theta_c$ and $\phi_c$ are the angles which define the direction of the difference vector $\vec{c}=\vec{p}_+-\vec{p}_-$.

The change of variables in the integral requires a reordering of steps starting from (\ref{eq:intermediate_dsigma}).
Rather than using spherical coordinates for $\vec{p}_\pm$ ($\vec{a}$ and $\vec{b}$), we replace $\dd^3a\dd^3b=-\frac{1}{8} \dd^3q \dd^3c$ and use spherical coordinates of $\vec{q}$, $\vec{c}$ i.e. $\dd^3q = q^2\dd q\dd\cos\theta_q \dd\phi_q$, $\dd^3c = c^2\dd c\dd\cos\theta_c \dd\phi_c$. Since $\vec{a}=\frac{1}{2}(\vec{q}+\vec{c})$ and $\vec{b}=\frac{1}{2}(\vec{q}-\vec{c})$ then the Jacobian $\left| \frac{\partial(\vec{a},\vec{b})}{\partial(\vec{q},\vec{c})}\right|=-\frac{1}{8}$.
We then replace $\dd q\dd c \dd\cos\theta_c = -8\frac{a^2b^2}{q^2c^2}\dd a \dd b \dd x$.
The Jacobian is calculated using
\begin{align}
	q^2 =& a^2 + b^2 + 2abx\;, \\
	c^2 =& a^2 + b^2 - 2abx\;, \\
	\cos \theta_c =& \frac{\sqrt{a^2-b^2}}{qc}\;,\\
	\left| \frac{\partial(q,c,\cos\theta_c)}{\partial(a,b,x)} \right| =& -8\frac{a^2b^2}{q^2c^2}\;.
\end{align}
Again $p_\pm\dd p_\pm=E_\pm \dd E_\pm$ and we use the $\delta$ function to integrate over $E_-$, then finally take $\frac{\omega}{2}\dd y=\dd E_+$.

\begin{table}[tbph]
	\centering
	\begin{tabular}{|c|c|c|}
		\hline
		$n$ & $v_n$ \\
		\hline
		\hline
		$1$ & $\frac{Q^{4}}{q^{4}}(\alpha\beta+abx-m^{2})$ \\ \hline
		$2$ & $-\frac{c}{\sqrt{2}}\sin\theta_c\cos\phi_c\left(\alpha-\beta-\frac{\omega}{q}c\cos\theta_c\right)$ \\ \hline
		$3$ & $-\frac{c}{\sqrt{2}}\sin\theta_c\cos\phi_c\left(\alpha-\beta-\frac{\omega}{q}c\cos\theta_c\right)$ \\ \hline
		$4$ & $-\frac{c^2}{4}\sin^2\theta_c+\alpha\beta-abx+m^2$ \\ \hline
		$5$ & $\frac{c^2}{2}\sin^2\theta_c\left(\cos^2\phi_c-\sin^2\phi_c\right)$ \\ \hline
		$6$ & $-c^2\sin^2\theta_c\cos\phi_c\sin\phi_c$ \\
		\hline
	\end{tabular}
	\caption{Summary of the kinematic prefactors with the $\theta_c$ and $\phi_c$ dependence explicit.}\label{tab:v_phic}
\end{table}

One can see from the expressions of the kinematic factors in \autoref{tab:v_phic} that an integral over $\phi_c$ results in zero for $v_2$, $v_3$, $v_5$ and $v_6$. The remaining $v_1$ and $v_4$ are independent of $\phi_c$ and will gain a factor of $2\pi$.
The $R_n$ are only dependent on $\theta_q$ (and $x$ via $q^2$). The integral over $\theta_q$ results in only the $K=0$ terms, as the Legendre polynomials integrate to zero except the zeroth (which gives $2$). There is no $\phi_q$ dependence which results in a further factor of $2\pi$.
The remaining variables can be expressed in terms of the magnitude of the outgoing momenta and the separation variable $\theta_{ee}$ (again $x=\cos\theta_{ee}$)
e.g. for completeness:
\begin{align}
	\sin\theta_c &= \frac{\sqrt{q^2c^2-(a^2-b^2)^2}}{qc} \;.
\end{align}

The resulting partial differential cross section after integration of $\theta_q$, $\phi_q$ and $\phi_c$ is
\begin{align}
	\frac{\dd^2 \sigma}{\dd y \dd x} = \frac{2\alpha^2\omega}{\pi}\frac{p_+p_-}{Q^4}\left[ v_1 \left( q^2 \textbf{E}_1^2 + q^4 \textbf{E}_2^2 \right) + v_4 \left( \omega^2 \textbf{E}_1^2 + q^2 \textbf{M}_1^2 + \frac{3}{4} \omega^2 q^2 \textbf{E}_2^2 \right) \right] \label{eq:intermediate_yr_integral}
\end{align}
using the shorthand
\begin{align}
	\textbf{E}_1^2 &= \frac{2\pi^2}{9k^2} \sum_{\ell_i s_i} \left| \mathcal{M}^{E1}_{0 s_i \ell_i 1} \right|^2 \;, \\
	\textbf{M}_1^2 &= \frac{2\pi^2}{9k^2} \sum_{\ell_i s_i} \left| \mathcal{M}^{M1}_{0 s_i \ell_i 1} \right|^2 \;, \\
    \textbf{E}_2^2 &= \frac{2\pi^2}{225k^2} \sum_{\ell_i s_i} \left| \mathcal{M}^{E2}_{0 s_i \ell_i 2} \right|^2 \;.
\end{align}
From the $p+\ce{^7Li}$ initial scattering state we include $J_i^{\pi_i} \in \{1^-, 1^+, 2^+\}$.  The prefactor is a combination of $\frac{1}{\hat{s}_P^2\hat{s}_T^2}=\frac{1}{8}$, the wavefunction normalization $\frac{4\pi}{k^2}$ and $\frac{4\pi}{[(2j+1)!!]^2}$ from the operators. In the low-energy approximation this last factor appears in the squared Coulomb operator as the factor $4\pi\frac{q^{2j}}{[(2j+1)!!]^2}$, in the squared electric $2\pi\frac{\omega^{2j}}{[(2j+1)!!]^2}\frac{j+1}{j}$ and in the magnetic as the same times $\frac{q^{2}}{\omega^{2}}$.

The kinematically allowed region of possible electron and positron energies is limited by the range $r-1 \leq y \leq 1-r$ where $r=\frac{2m}{\omega}$.
Using $y$ and $r$, the parameters may all be expressed with the dimensionfull $\omega$ factored out, i.e.
\begin{align}
	E_{\pm} &= \frac{\omega}{2}\left(1\pm y\right)\;, \\
	p_\pm &= \frac{\omega}{2}\sqrt{(1\pm y)^2 -r^2} \;, \\
	E_+E_- &= \alpha\beta = \frac{\omega^2}{4}\left(1-y^2\right) \;.
\end{align}
For simplification of the expressions, we define:
\begin{align}
	s &= \sqrt{(1+y)^2-r^2}\sqrt{(1-y)^2-r^2} \;, \\
	t &= y^2 - r^2\;.
\end{align}
We then have:
\begin{align}
	a^2+b^2 &= \frac{\omega^2}{2}\left(1+t\right)\;, \\
	ab &= \frac{\omega^2}{4}s\;,
\end{align}
and
\begin{align}
	q^2 &= \frac{\omega^2}{2}\left( 1+t+sx \right)\;, \\
	Q^2 &= \frac{\omega^2}{2}( 1- t -sx)\;, \\
	c^2 &= \frac{\omega^2}{2}\left(1+t-sx\right)\;.
\end{align}
The partial differential cross section is then
\begin{align}
	\frac{\dd^2 \sigma}{\dd y\dd x} 	=& \frac{\alpha^2\omega^3}{2\pi} \bigg[ \frac{s(1-y^2-r^2+sx)}{2(1+t+sx)} \left( \textbf{E}_1^2 + \frac{\omega^2}{2}(1+t+sx) \textbf{E}_2^2 \right) \nonumber \\ 
	& + \frac{s}{(1-t-sx)^2} \left( 1- t -\frac{s^2(1+x^2) + 2(1+t)sx }{2(1+t+sx)}\right) \bigg[ \textbf{E}_1^2 + \frac{1}{2}(1+t+sx)\left( \textbf{M}_1^2 + \frac{3}{4}\omega^2\textbf{E}_2^2 \right) \bigg] \bigg]\;. \label{eq:partial_ts}
\end{align}
This result can be integrated numerically with quadrature methods, i.e. the \texttt{scipy.dblquad} Python function or in \texttt{Mathematica} with the nuclear transition matrix elements for each value of $\omega$ calculated with the NCSMC.

\subsection{Comparison to Pair Production in Bound State Decay}
\label{subsec:pairprod_bound_appendix}

\begin{figure}[tbph]
	\begin{center}
		\includegraphics{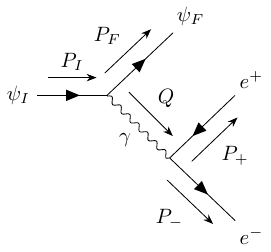}
	\end{center}\caption{The Feynman diagram for the pair production process.}
	\label{fig:feyn_pairprod}
\end{figure}

The process of pair production is simpler to calculate in a bound-bound transition,
just as $\gamma$-decay is simpler than radiative capture.

We calculate the decay rate of a $1\to3$ process, i.e.
\begin{align}
	\dd\Gamma= & \frac{1}{2M_I}\bar{\sum_{i}}\sum_{f}\sum_{s_{+}s_{-}}\left|\mathcal{M}_{FI}^{s_{+}s_{-}}\right|^{2}\frac{\dd^{3}p_{+}}{(2\pi)^{3}2E_{+}}\frac{\dd^{3}p_{-}}{(2\pi)^{3}2E_{-}}\frac{\dd^{3}p_{F}}{(2\pi)^{3}2E_{F}}\;,
\end{align}
In this case, the squared amplitude (summed over lepton polarizations) is:
\begin{align}
	\sum_{s_+s_-}\left|\mathcal{M}^{s_+s_-}_{FI}\right|^2 &= \ell^{\mu\nu}\mathcal{J}_\mu\mathcal{J}_\nu^* (2\pi)^4 \delta(P_I-P_F-P_+-P_-) 4M_IE_F \;.
\end{align}

Here, rather than replacing $\mathcal{J}_z\to\frac{\omega}{q}\mathcal{\rho}$, we equivalently replace $\rho \to -\frac{Q^2}{q^2}\rho$ and set $\mathcal{J}_z=0$, then $\vec{\mathcal{J}}$ becomes $\vec{\mathcal{J}}^T$, 
where $\vec{\mathcal{J}}^{T}$ is the
transverse current:
\begin{equation}
	\vec{\mathcal{J}}^{T}=\sum_{\lambda=\pm 1}\left(\vec{e}^*_{\lambda}\cdot\vec{\mathcal{J}}\right)\vec{e}_\lambda\label{eq:transverse_J}\;,
\end{equation}
or equivalently 
\begin{equation}
	\vec{\mathcal{J}}^T=\left(\vec{e}_x\cdot \vec{\mathcal{J}}\right)\vec{e}_x + \left(\vec{e}_y\cdot\vec{\mathcal{J}}\right) \vec{e}_y \;.
\end{equation}

Defining $\varrho=\frac{-Q^{2}}{q^{2}}\rho$, we have the
intermediate result:
\begin{align}
	\frac{1}{4}\ell^{\mu\nu}\mathcal{J}_{\mu}^{*}\mathcal{J}_{\nu}= & 2E_{+}E_{-}\left|\varrho\right|^{2}\nonumber \\
	&-E_{+}\varrho\left(\vec{p}_{-}\cdot\vec{\mathcal{J}}^{T*}\right)-E_{-}\varrho^{*}\left(\vec{p}_{+}\cdot\vec{\mathcal{J}}^{T}\right)-E_{+}\varrho^{*}\left(\vec{p}_{-}\cdot\vec{\mathcal{J}}^{T}\right)-E_{-}\varrho\left(\vec{p}_{+}\cdot\vec{\mathcal{J}}^{T*}\right)\nonumber \\
	& +\left(\vec{p}_{+}\cdot\vec{\mathcal{J}}^{T}\right)\left(\vec{p}_{-}\cdot\vec{\mathcal{J}}^{T*}\right)+\left(\vec{p}_{+}\cdot\vec{\mathcal{J}}^{T*}\right)\left(\vec{p}_{-}\cdot\vec{\mathcal{J}}^{T}\right)\nonumber \\
	&-\left(\left|\varrho\right|^{2}-\left|\mathcal{J}^{T}\right|^{2}\right)\left(\bar{E}E-\vec{p}_{+}\cdot\vec{p}_{-}-m^{2}\right)\;.
\end{align}
Since we evaluate matrix elements between $\ket{i}=\ket{J_iM_i}$ and $\ket{f}=\ket{J_fM_f}$ which have fixed $M$, then
the terms which mix $\varrho$ and $\vec{\mathcal{J}}^{T}$ must vanish
as $\rho$ does not change the $M$ of the initial state but $\vec{\mathcal{J}}^{T}$
does (as $\vec{\mathcal{J}}^T$ consists of the operators $\mathcal{J}_\pm$). The terms which mix $\lambda=1$ and $\lambda=-1$ will also be zero e.g. $\mathcal{J}_+ \mathcal{J}_-^*$.

This is equivalent to removing all the terms except $n=1$ and $n=4$ in (\ref{eq:full_diff_sigma}) as $v_{4}$ exactly matches the coefficient in front of
$\left|\mathcal{M}^{EM}\right|^{2}$, since
\begin{align*}
  c_{x}^{2}+c_{y}^{2}
  =& \left|\vec{c}-\frac{\vec{c}\cdot\vec{q}}{q^{2}}\vec{q}\right|^{2}\;, \\
  \alpha\beta-\vec{a}\cdot\vec{b}+m^{2}-\frac{c_{x}^{2}+c_{y}^{2}}{4} =&
  \alpha\beta+m^{2}-\frac{(\vec{a}\cdot\vec{q})(\vec{b}\cdot\vec{q})}{q^{2}}\;.
\end{align*}
The lepton tensor and nuclear current contractions result in
\begin{align}
	\frac{1}{4}\ell^{\mu\nu}\mathcal{J}_{\mu}\mathcal{J}_{\nu}^{*}
	= & \frac{Q^{4}}{q^{4}}\left|\rho\right|^{2}\left(E_{+}E_{-}+\vec{p}_{+}\cdot\vec{p}_{-}-m^{2}\right)+\left|\mathcal{J}^{T}\right|^{2}\left(E_{+}E_{-}+m^{2}-\frac{\left(\vec{p}_{+}\cdot\vec{q}\right)\left(\vec{p}_{-}\cdot\vec{q}\right)}{q^{2}}\right)\label{eq:lJJ}\;.
\end{align}
where the charge density relates directly to the Coulomb operators, i.e.
\begin{align}
	\bar{\sum_{i}}\sum_{f}\left|\rho^{2}\right|= & \frac{4\pi}{\hat{J}_{i}^{2}}\sum_{J\geq0}\left|\mathcal{C}_{J}\right|^{2}\equiv\left|\mathcal{M}^{C}\right|^{2}\;,
\end{align}
while the transverse current relates to the transverse electric and
magnetic operators

\begin{align}
	\bar{\sum_{i}}\sum_{f}\left|\mathcal{J}^{T}\right|^{2}= & \frac{4\pi}{\hat{J}_{i}^{2}}\sum_{J\geq1}\left|\mathcal{T}_{J}^{E}\right|^{2}+\left|\mathcal{T}_{J}^{M}\right|^{2}\equiv\left|\mathcal{M}^{EM}\right|^{2}\;.
\end{align}

Putting everything together, we first do the integral over the final nucleus momentum:
\begin{align}
	\dd\Gamma= & \left(\frac{4\pi\alpha}{Q^{2}}\right)^{2}\frac{\dd^{3}p_+}{(2\pi)^{3}2E_+}\frac{\dd^{3}p_-}{(2\pi)^{3}2E_-}2\pi\delta(M_I-E_F-E_+-E_-)\bar{\sum_{i}}\sum_{f}\ell^{\mu\nu}\mathcal{J}_{\mu}^{*}\mathcal{J}_{\nu}\;.
\end{align}

Second, we do the angular integrals. There are four angles determining the directions of the positron and electron momenta. The amplitude only depends on the relative separation angle $\theta_{ee}$. Therefore by measuring the direction of $\vec{p}_+$ relative to $\vec{p}_-$ we can replace the integration $\dd \cos\theta_+ \to \dd  \cos\theta_{ee} = \dd x$. The remaining three angles ($\dd\phi_+ \dd\cos\theta_- \dd\phi_-$) then give a constant factor ($8\pi^2$) in the total rate, i.e.
\begin{align}
	\dd^3 p_+ \dd^3 p_- = 2\pi \dd x p_+^2 \dd p_+ 4\pi p_-^2 \dd p_- \;.
\end{align}

Third, we insert the lepton-nuclear tensor contraction and do the integral
over $\dd p_-$ using the remaining $\delta$. We get a factor of $\frac{\sqrt{p_-^{2}+m^{2}}}{p_-}=\frac{E_-}{p_-}$
and a recoil factor $f_{r}=\left(1+\frac{E_{-}}{E_{+}}\left(1+\frac{p_{+}x}{p_{-}}\right)\right)^{-1}$.
We can change integration variables $p_{+}\dd p_{+}=E_{+}\dd E_{+}$, and further $\frac{\omega}{2}\dd y=\dd E_{+}$ using (\ref{eq:y}),
resulting in the differential decay rate:
\begin{align}
	\frac{\dd\Gamma}{\dd x\dd y}= & \frac{2\alpha^{2}\omega}{\pi Q^{4}}p_{-}p_{+}f_{R}\bigg[\frac{Q^{4}}{q^{4}}\left|\mathcal{M}^{C}\right|^{2}\left(E_{+}E_{-}+p_{+}p_{-}x-m^{2}\right)\nonumber\\
	&+\left|\mathcal{M}^{EM}\right|^{2}\left(E_{+}E_{-}+m^{2}-\frac{\left(\vec{p}_{+}\cdot\vec{q}\right)\left(\vec{p}_{-}\cdot\vec{q}\right)}{q^{2}}\right)\bigg]\label{eq:dGamma_pp}\;.
\end{align}
This matches the formula of \cite{Hayes2022}.

In order to evaluate this rate using NCSMC transition matrix elements, we take the diagonal matrix elements e.g. $\sum_{\ell_is_iJ_i}\left|\mathcal{C}_{J_f\ell_i s_i J_i}\right|^2$, with the appropriate normalization factors, equivalent to the $K=0$ terms of (\ref{eq:NN}). The result is proportional to the partially integrated differential cross section of (\ref{eq:intermediate_yr_integral}) and (\ref{eq:partial_ts}).
\end{widetext}


%

\end{document}